\documentclass[aps,prd,twocolumn,superscriptaddress,nofootinbib,longbibliography,floatfix]{revtex4-2}
\usepackage{amsmath,amssymb,graphicx,bm,xcolor,textcomp}
\usepackage{tikz}
\usetikzlibrary{decorations.pathmorphing}
\usepackage[colorlinks=true,linkcolor=blue,citecolor=blue,urlcolor=blue]{hyperref}

\def\beq{\begin{equation}}
\def\eeq{\end{equation}}
\def\bea{\begin{eqnarray}}
\def\eea{\end{eqnarray}}

\newcommand{\Lb}{\left(}
\newcommand{\Rb}{\right)}

\newcommand{\eq}[1]{Eq.~(\ref{#1})}

\newcommand{\cev}[1]{\reflectbox{\ensuremath{\vec{\reflectbox{\ensuremath{#1}}}}}}
\newcommand{\bas}{\bar\alpha_s}
\newcommand{\un}[1]{\underline{#1}}

\newcommand{\tr}{\mathrm{tr}}

\begin{document}

\title{Small-$x$ helicity evolution with gluon saturation}

\author{M. Roa}
\email{michael.roa@upla.cl}
\affiliation{Departamento de F\'isica, Universidad T\'ecnica Federico Santa Mar\'ia, Casilla 110-V, Valpara\'iso, Chile}
\affiliation{Facultad de Ingenier\'ia, Laboratorio DataScience, Universidad de Playa Ancha, Leopoldo Carvallo 270, Valpara\'iso, Chile}

\begin{abstract}
We include gluon saturation in the revised small-$x$ helicity evolution at large $N_c$. Keeping the unpolarized dipole $S$-matrices, which are set to one in the double-logarithmic approximation, we derive the evolution equations for the polarized dipole amplitudes $G$ and $G_2$ and their neighbor amplitudes. The equations reduce exactly to the known ones for $S=1$ and contain new terms proportional to $1-S$, including one in the equation for $G_2$ that arises because the real and virtual contributions that cancel each other for $S=1$ are weighted differently by the $S$-matrix of the small daughter dipole; this term is ultraviolet finite and acts only when the daughter dipole is saturated. Inside the saturation region the growth rate of $G$ drops from $3.66\sqrt{\alpha_sN_c/2\pi}$ to $2\sqrt{1-\lambda}\,\sqrt{\alpha_sN_c/2\pi}$, with $\lambda$ the growth exponent of the saturation scale, a result that we derive exactly for $\lambda\to0$ and $\lambda=1$ and with a WKB analysis in between, while $G_2$ oscillates without growing. In the asymptotic regime of the equations, the gluon helicity $\Delta G$ reaches a maximum and decreases, while the quark helicity $\Delta\Sigma$ and the $g_1$ structure function keep growing with an exponent set by the dipoles at the edge of the saturation region, which we obtain in closed form; saturation raises the value of $\alpha_s$ below which the integral of $\Delta\Sigma$ converges from $0.156$ to $0.21$ for $\lambda=0.25$. In the experimental range $10^{-2}\ge x\ge10^{-6}$ we solve the equations numerically with a target described by the running-coupling Balitsky--Kovchegov equation, whose saturation scale is close to that extracted from the HERA data. At the reach of the Electron-Ion Collider, $x\sim10^{-4}$, and $Q^2=10$~GeV$^2$, saturation reduces $\Delta G$ by a factor of about three and $\Delta\Sigma$ by a factor of about two; both distributions still grow there, with reduced exponents. We compare three choices of the starting point of the evolution, $x_0=0.1$ and $0.01$, and show that the dependence on this choice is lost toward $x=10^{-6}$.
\end{abstract}

\maketitle

\section{Introduction}
\label{sec:intro}

How the spin of the proton is distributed among its quarks and gluons remains one of the central open questions of QCD. The EMC measurement of the spin asymmetry in polarized deep inelastic scattering (DIS) showed that the quark spin accounts for only a small fraction of the proton spin~\cite{EuropeanMuon:1987isl,EuropeanMuon:1989yki}, and three decades of experimental and theoretical work have followed (see Refs.~\cite{Aidala:2012mv,Leader:2013jra,Ji:2020ena} for reviews). In the light-cone gauge the spin of a longitudinally polarized proton decomposes according to the Jaffe--Manohar sum rule~\cite{Jaffe:1989jz},
\beq
\label{eq:JM}
\frac{1}{2} = \frac{1}{2}\,\Delta\Sigma\Lb Q^2\Rb + \Delta G\Lb Q^2\Rb + L_q\Lb Q^2\Rb + L_g\Lb Q^2\Rb,
\eeq
where $L_q$ and $L_g$ are the quark and gluon orbital angular momenta (OAM) and
\beq
\label{eq:moments}
\begin{split}
\Delta\Sigma\Lb Q^2\Rb &= \int\limits_0^1 dx\, \Delta\Sigma\Lb x,Q^2\Rb, \\
\Delta G\Lb Q^2\Rb &= \int\limits_0^1 dx\, \Delta G\Lb x,Q^2\Rb
\end{split}
\eeq
are the first moments of the flavor-singlet quark helicity distribution and of the gluon helicity distribution. A gauge-invariant alternative decomposition was proposed by Ji~\cite{Ji:1996ek}; the relation between the two is discussed in Ref.~\cite{Leader:2013jra}. Both decompositions share the helicity terms in \eq{eq:moments}, which are the subject of this paper.

Global analyses of polarized DIS, semi-inclusive DIS and polarized proton--proton data~\cite{deFlorian:2009vb,deFlorian:2014yva,Nocera:2014gqa,Ethier:2017zbq} have confirmed that the quark helicity accounts for only part of the proton spin and, after the high-statistics RHIC jet data~\cite{STAR:2014wox}, have found evidence for a nonvanishing gluon polarization in the range of $x$ probed by the data~\cite{deFlorian:2014yva}. Information from low momentum fractions, however, is lacking~\cite{deFlorian:2014yva}. The integrals in \eq{eq:moments} extend down to $x=0$, and their small-$x$ part is controlled by theory rather than by data. The future Electron-Ion Collider (EIC) will extend the polarized DIS data to considerably smaller $x$~\cite{Accardi:2012qut,AbdulKhalek:2021gbh}. A reliable theoretical understanding of the helicity distributions at small $x$ is needed both to interpret those data and to estimate the amount of spin that remains at values of $x$ beyond their reach.

The small-$x$ asymptotics of helicity is governed by double logarithms. Unlike the unpolarized case, where the leading high-energy evolution resums powers of $\alpha_s \ln\Lb 1/x\Rb$, the helicity distributions receive contributions enhanced by $\alpha_s \ln^2\Lb 1/x\Rb$~\cite{Kirschner:1983di}. Building on the infrared evolution equation approach of Ref.~\cite{Kirschner:1983di}, the double logarithms of the $g_1$ structure function were resummed by Ermolaev, Manaenkov and Ryskin~\cite{Ermolaev:1995fx} and by Bartels, Ermolaev and Ryskin (BER)~\cite{Bartels:1995iu,Bartels:1996wc}; see also Ref.~\cite{Ermolaev:2003zx} for running coupling effects. The double-logarithmic approximation (DLA) goes beyond the fixed-order DGLAP evolution~\cite{Altarelli:1977zs}, whose polarized splitting functions are known to three loops~\cite{Moch:2014sna}.

A different approach was developed by Kovchegov, Pitonyak and Sievert (KPS)~\cite{Kovchegov:2015pbl,Kovchegov:2016zex,Kovchegov:2016weo,Kovchegov:2017jxc,Kovchegov:2017lsr,Kovchegov:2018znm}, based on the $s$-channel (shock wave) formalism of small-$x$ physics~\cite{Mueller:1994jq,Balitsky:1995ub,Kovchegov:1999yj,Kovchegov:2012mbw}. In this framework the helicity distributions are expressed through polarized dipole amplitudes, which are expectation values of Wilson-line operators with one sub-eikonal, helicity-dependent insertion. Their evolution equations are closed in the limit of a large number of colors $N_c$, and in the large-$N_c\,\&\,N_f$ limit, with $N_f$ the number of quark flavors. Throughout this paper we use
\beq
\label{eq:abar}
\bas \equiv \frac{\alpha_s N_c}{2\pi}.
\eeq
The first versions of these equations gave the quark helicity intercept $\alpha_h^q = \Lb 4/\sqrt{3}\Rb\sqrt{\bas}\approx 2.31\sqrt{\bas}$~\cite{Kovchegov:2016weo,Kovchegov:2017jxc} and a gluon helicity intercept $\approx 1.88\sqrt{\bas}$~\cite{Kovchegov:2017lsr}. Later, Cougoulic, Kovchegov, Tarasov and Tawabutr (KPS-CTT) found that the mixing with an additional sub-eikonal operator, $\cev{D}^iD^i$, was missing in the earlier works, and derived the revised evolution equations, which involve a new polarized dipole amplitude $G_2$~\cite{Cougoulic:2022gbk}. The numerical solution of the revised large-$N_c$ equations gave~\cite{Cougoulic:2022gbk}
\beq
\label{eq:asymp_intro}
\Delta\Sigma\Lb x,Q^2\Rb \sim \Delta G\Lb x,Q^2\Rb \sim g_1\Lb x,Q^2\Rb \sim \left(\frac{1}{x}\right)^{\alpha_h},
\eeq
with $\alpha_h \approx 3.66\sqrt{\bas}$, in agreement with BER at that precision. Borden and Kovchegov then solved these equations exactly and obtained the intercept in analytic form~\cite{Borden:2023ugd},
\beq
\label{eq:alphah}
\begin{split}
\alpha_h &= \frac{4}{3^{1/3}} \sqrt{\mathrm{Re}\left[\left(-9+i\sqrt{111}\right)^{1/3}\right]}\;\sqrt{\bas} \\
&\approx 3.66074\,\sqrt{\bas},
\end{split}
\eeq
which differs from the BER value $3.66394\sqrt{\bas}$ beyond the second decimal place, together with a resummed polarized anomalous dimension that agrees with the existing fixed-order results~\cite{Borden:2023ugd}. The formalism has since been extended to single-logarithmic accuracy and running coupling~\cite{Kovchegov:2021lvz}, to include the quark-to-gluon and gluon-to-quark transition operators in the large-$N_c\,\&\,N_f$ limit~\cite{Borden:2024bxa}, and to the OAM distributions~\cite{Kovchegov:2023yzd}. In the large-$N_c\,\&\,N_f$ limit the quark contributions make the helicity distributions oscillate as functions of $\ln\Lb 1/x\Rb$~\cite{Kovchegov:2020hgb}. Related sub-eikonal studies of small-$x$ scattering can be found in Refs.~\cite{Chirilli:2018kkw,Altinoluk:2020oyd,Kovchegov:2021iyc}, and the spin and orbital angular momentum at small $x$ have also been studied in other frameworks~\cite{Hatta:2016aoc,Boussarie:2019icw}. On the phenomenological side, the small-$x$ helicity evolution has been used to describe the world polarized DIS and SIDIS data at $x<0.1$~\cite{Adamiak:2021ppq,Adamiak:2023yhz}.

Equation~(\ref{eq:asymp_intro}) has a striking consequence for the spin sum rule. The integral $\int_0 dx\,\Lb 1/x\Rb^{\alpha_h}$ converges at small $x$ only for $\alpha_h<1$. With $N_c=3$, \eq{eq:alphah} gives $\alpha_h<1$ only for
\beq
\label{eq:threshold0}
\alpha_s < \frac{2\pi}{N_c\,\Lb 3.66074\Rb^2} \approx 0.156 .
\eeq
For realistic values $\alpha_s \approx 0.2$--$0.3$ one finds $\alpha_h \approx 1.1$--$1.4$, and the moments~(\ref{eq:moments}), if computed with the DLA asymptotics~(\ref{eq:asymp_intro}) all the way down to $x=0$, would diverge. Such a power-law growth cannot continue indefinitely: at sufficiently small $x$ the gluon density of the target saturates~\cite{Gribov:1983ivg,Mueller:1985wy}, and the growth of the observables is expected to slow down.

Gluon saturation is described by the color glass condensate (CGC)~\cite{McLerran:1993ni,McLerran:1993ka,Gelis:2010nm,Weigert:2005us,Jalilian-Marian:2005ccm,Kovchegov:2012mbw}. The high-energy evolution of the unpolarized dipole $S$-matrix is given by the Balitsky--Kovchegov (BK) equation~\cite{Balitsky:1995ub,Kovchegov:1999yj,Kovchegov:1999ua} and, beyond the large-$N_c$ limit, by the JIMWLK equations~\cite{Jalilian-Marian:1997qno,Jalilian-Marian:1997jhx,Iancu:2000hn,Ferreiro:2001qy}. The nonlinear term generates the saturation scale $Q_s\Lb x\Rb$, which grows as a power of $1/x$~\cite{Mueller:2002zm}, and leads to geometric scaling~\cite{Stasto:2000er,Iancu:2002tr}. Deep inside the saturation region the dipole $S$-matrix decreases as a Gaussian in $\ln\Lb Q_s^2r^2\Rb$, as found by Levin and Tuchin~\cite{Levin:1999mw,Levin:2001cv}. With running coupling corrections~\cite{Balitsky:2006wa,Kovchegov:2006vj,Albacete:2007yr}, the BK equation describes the small-$x$ HERA data~\cite{H1:2015ubc,Albacete:2009fh,Albacete:2010sy,Lappi:2013zma}, as do saturation models such as the Golec-Biernat--W\"usthoff (GBW) model~\cite{Golec-Biernat:1998zce,Golec-Biernat:1999qor} and its extensions~\cite{Bartels:2002cj,Iancu:2003ge,Kowalski:2003hm,Rezaeian:2012ji}. Corrections beyond leading logarithmic accuracy have also been computed~\cite{Balitsky:2007feb,Beuf:2014uia,Iancu:2015vea,Ducloue:2019ezk}. For reviews of the phenomenology see Refs.~\cite{Albacete:2014fwa,Morreale:2021pnn}.

Saturation enters the helicity evolution in a natural way. In the derivation of the KPS-CTT equations, the emission of a soft gluon by a polarized dipole involves the unpolarized dipole $S$-matrices $S_{20}$ and $S_{21}$ of the daughter dipoles~\cite{Cougoulic:2022gbk}. Since the evolution of $S$ is single-logarithmic, the DLA equations are obtained by setting $S_{20}=S_{21}=1$~\cite{Cougoulic:2022gbk}. This is consistent at strict DLA accuracy, and it is exact in the dilute regime $Q_s^2x_{10}^2\ll 1$. It cannot be correct, however, for dipoles larger than $1/Q_s$, where $S\to 0$. The question of how to combine the DLA helicity evolution with the single-logarithmic unpolarized evolution was already discussed in Ref.~\cite{Kovchegov:2015pbl}, where the equations were written with the $S$-matrices kept, and it was argued that such a combination is likely justified, in analogy with the earlier treatment of the DLA Reggeon evolution with saturation by Itakura, Kovchegov, McLerran and Teaney~\cite{Itakura:2003jp}. In that problem saturation was found to reduce the Reggeon intercept~\cite{Itakura:2003jp}. Saturation effects in helicity have also been studied recently from other angles. Li evaluated the saturation-induced helicity-dependent gluon field in a quasi-classical framework and its effect on the double-spin asymmetry in diffractive dijet production~\cite{Li:2024xra}, and derived a nonlinear evolution equation for the sub-eikonal dipole gluon distribution at single-logarithmic accuracy, which obeys the Levin--Tuchin law in the saturation regime~\cite{Li:2026azt}. Brown, DeBrizzi and Kovner showed that for strong fields the naive eikonal expansion mixes eikonal orders, and proposed a way to rectify this~\cite{Brown:2026qlc}. To the best of our knowledge, however, the revised large-$N_c$ helicity evolution of Ref.~\cite{Cougoulic:2022gbk} has not been solved with the unpolarized $S$-matrices kept, and the effect of saturation on the intercept~(\ref{eq:alphah}) and on the convergence of the moments~(\ref{eq:moments}) is not known.

In this paper we address this problem. Our main results are the following.
\begin{enumerate}
\item We derive the DLA evolution equations for the polarized dipole amplitudes $G$ and $G_2$ and their neighbor amplitudes $\Gamma$ and $\Gamma_2$ at large $N_c$, keeping the unpolarized $S$-matrices. The equation for $G$ acquires terms proportional to $S$ and to $1-S$. In the equation for $G_2$ the real and virtual ultraviolet contributions, which cancel each other for $S=1$, are weighted differently by the $S$-matrix of the small daughter dipole; for $S\neq1$ they leave a new, ultraviolet-finite term proportional to $\Lb 1-S\Rb\,\Gamma_2$, which acts only when the daughter dipole is saturated. For $S=1$ our equations reduce exactly to those of Ref.~\cite{Cougoulic:2022gbk}, and our numerical solution reproduces the intercept~(\ref{eq:alphah}).
\item Inside the saturation region the equations simplify and can be solved analytically. The polarized amplitude $G$ keeps growing, but with the reduced intercept $2\sqrt{1-\lambda}\,\sqrt{\bas}$, where $\lambda$ is the growth exponent of the saturation scale, $Q_s^2\propto \Lb 1/x\Rb^{\lambda}$. In the limit $\lambda\to0$ the intercept $2\sqrt{\bas}$ is exact. The amplitude $G_2$ stops growing and oscillates.
\item In the asymptotic regime of the equations, the gluon helicity distribution $\Delta G\Lb x,Q^2\Rb$ reaches a maximum and then decreases, so that its contribution to \eq{eq:JM} is finite. The quark helicity $\Delta\Sigma$ and the structure function $g_1$ keep growing, with an exponent set by the dipoles at the edge of the saturation region. We obtain this exponent in closed form, $\alpha_{\Delta\Sigma} = \phi_0\Lb \lambda\Rb\sqrt{\bas}$, from a saddle point of the exact solution of Ref.~\cite{Borden:2023ugd}, with $\phi_0\Lb 0\Rb=3.66074$ and $\phi_0\Lb 0.25\Rb\approx 3.18$. As a consequence, saturation raises the value of $\alpha_s$ below which the integral of $\Delta\Sigma$ converges, from $0.156$ in \eq{eq:threshold0} to $\alpha_s^\ast\Lb \lambda\Rb = 2\pi/\Lb N_c\,\phi_0^2\Lb \lambda\Rb\Rb$, for example $\alpha_s^\ast \approx 0.21$ for $\lambda=0.25$.
\item In the experimental range $10^{-2}\ge x\ge10^{-6}$ we solve the equations numerically with $S$ given by the running-coupling BK equation, for a target whose saturation scale is close to that of the proton at HERA. The asymptotic regime is not reached in this range. Saturation suppresses both distributions, $\Delta G$ more strongly: at the EIC reach, $x\sim10^{-4}$, and $Q^2=10$~GeV$^2$, $\Delta G$ is reduced to about $30\%$ and $\Delta\Sigma$ to about $60\%$ of their values without saturation, and the effective exponents are reduced from $1.15$ to $0.77$ and from $1.16$ to $0.99$ for $\alpha_s=0.25$. We compare three choices of the starting points of the helicity and target evolutions, $x_0=0.1$ as in the polarized analyses of Refs.~\cite{Adamiak:2021ppq,Adamiak:2023yhz} and $x_0=0.01$ as in the rcBK descriptions of HERA~\cite{Albacete:2009fh,Albacete:2010sy}, and find that the dependence on this choice disappears toward $x=10^{-6}$.
\end{enumerate}
This paper is theoretical: we study the structure of the evolution, the asymptotics of its solutions, and the size of the saturation effects in the experimental range, but we do not attempt a fit to data.

The paper is organized as follows. In Sec.~\ref{sec:S} we recall the helicity evolution equations with the unpolarized $S$-matrices and the counting of logarithms. In Sec.~\ref{sec:DLA} we derive the DLA equations with saturation and discuss their properties. Section~\ref{sec:asympt} is devoted to the asymptotic regime: the analytic solution inside the saturation region, the closed formula for the exponent of $\Delta\Sigma$ and the convergence of the helicity integrals. In Sec.~\ref{sec:numerics} we present the numerical solution in the experimental range of $x$ with the running-coupling BK equation, and the helicity distributions $\Delta G$ and $\Delta\Sigma$ for three choices of the starting point of the evolution. We discuss the limitations of our approach in Sec.~\ref{sec:discussion} and conclude in Sec.~\ref{sec:conclusions}. The derivation of the equations for the neighbor amplitudes, the angular averages of all the kernels, the numerical method and the small-$c$ expansion of the saddle point are presented in Appendices~\ref{app:neighbor}, \ref{app:UV}, \ref{app:numerics} and~\ref{app:saddle}.

\section{Helicity evolution with unpolarized $S$-matrices}
\label{sec:S}

In this section we collect the ingredients of the KPS-CTT formalism~\cite{Cougoulic:2022gbk} that we need, and we identify where the unpolarized dipole $S$-matrices enter. We follow the notation of Ref.~\cite{Cougoulic:2022gbk} throughout: ${\un x}_{ij}={\un x}_i-{\un x}_j$ is the separation between the transverse positions of the partons $i$ and $j$, $x_{ij}=|{\un x}_{ij}|$, $z$ is the smallest light-cone momentum fraction of the softest parton in the projectile dipole, $s$ is the center-of-mass energy squared and $\Lambda$ is an infrared cutoff.

\subsection{Polarized dipole amplitudes and observables}
\label{sec:amplitudes}

In the shock-wave formalism the target is described by light-cone Wilson lines $V_{\un x}$ in the fundamental representation and $U_{\un x}$ in the adjoint representation~\cite{Balitsky:1995ub,Kovchegov:2012mbw}. The unpolarized dipole $S$-matrix is
\beq
\label{eq:Sdef}
S_{10}\Lb zs\Rb = \frac{1}{N_c}\left\langle \mathrm{T}\,\tr\left[V_{\un 0}V_{\un 1}^\dagger\right]\right\rangle\Lb zs\Rb,
\eeq
where $\mathrm{T}$ denotes time ordering of the operators and the angle brackets denote the average over the target state~\cite{Kovchegov:2015pbl,Cougoulic:2022gbk}. The $S$-matrix is assumed to be real, i.e., the odderon contribution is neglected as in Ref.~\cite{Cougoulic:2022gbk}. Following Ref.~\cite{Cougoulic:2022gbk}, we use two such insertions. The first one contains the gluon field strength $F^{12}$ and the quark axial current $\bar\psi\gamma^+\gamma^5\psi$; it defines the polarized Wilson line $V^{\mathrm{pol}[1]}_{\un x}$ and the polarized dipole amplitude of the first kind,
\beq
\label{eq:Qdef}
\begin{split}
Q_{10}\Lb zs\Rb \equiv \frac{1}{2N_c}\,\mathrm{Re}\,\Big\langle\!\Big\langle & \mathrm{T}\,\tr\left[V_{\un 0}V_{\un 1}^{\mathrm{pol}[1]\dagger}\right] \\
&+ \mathrm{T}\,\tr\left[V_{\un 1}^{\mathrm{pol}[1]}V_{\un 0}^\dagger\right]\Big\rangle\!\Big\rangle\Lb zs\Rb,
\end{split}
\eeq
where the double angle brackets denote the target average with the energy suppression factor scaled out~\cite{Cougoulic:2022gbk}. Its gluon part is the amplitude $G_{10}\Lb zs\Rb$, defined in the same way with $V^{\mathrm{G}[1]}$ instead of $V^{\mathrm{pol}[1]}$; at large $N_c$ the two are not distinguished and one sets $Q\approx G$~\cite{Cougoulic:2022gbk,Kovchegov:2020hgb}. The second insertion contains the operator $D^i-\cev{D}^i$, which is related to the Jaffe--Manohar polarized gluon distribution, and defines the polarized dipole amplitude of the second kind~\cite{Cougoulic:2022gbk,Kovchegov:2017lsr}
\beq
\label{eq:Gidef}
G^i_{10}\Lb zs\Rb \equiv \frac{1}{2N_c}\Big\langle\!\Big\langle \tr\Big[V_{\un 0}^\dagger V_{\un 1}^{i\,\mathrm{G}[2]} + \big(V_{\un 1}^{i\,\mathrm{G}[2]}\big)^\dagger V_{\un 0}\Big]\Big\rangle\!\Big\rangle\Lb zs\Rb.
\eeq
After integration over the impact parameter, $G^i$ is decomposed as~\cite{Kovchegov:2017lsr,Cougoulic:2022gbk}
\beq
\label{eq:decomp}
\begin{split}
\int d^2\Big(\frac{x_1+x_0}{2}\Big)\,G^i_{10}\Lb zs\Rb ={}& x_{10}^i\,G_1\Lb x_{10}^2,zs\Rb \\
&+ \epsilon^{ij}x_{10}^j\,G_2\Lb x_{10}^2,zs\Rb,
\end{split}
\eeq
and only $G_2$ contributes to the helicity observables.

The flavor-singlet quark helicity PDF, the gluon helicity PDF and the $g_1$ structure function are given in terms of these amplitudes by~\cite{Cougoulic:2022gbk}
\begin{widetext}
\begin{align}
\label{eq:DSigma}
\Delta\Sigma\Lb x,Q^2\Rb &= -\frac{N_cN_f}{2\pi^3}\int\limits_{\Lambda^2/s}^1\frac{dz}{z}\!\!\int\limits_{\frac{1}{zs}}^{\min\left\{\frac{1}{zQ^2},\frac{1}{\Lambda^2}\right\}}\!\!\frac{dx_{10}^2}{x_{10}^2}\,\big[Q+2G_2\big]\Lb x_{10}^2,zs\Rb, \\
\label{eq:DG}
\Delta G\Lb x,Q^2\Rb &= \frac{2N_c}{\alpha_s\pi^2}\left[\Big(1+x_{10}^2\frac{\partial}{\partial x_{10}^2}\Big)G_2\Big(x_{10}^2,zs=\frac{Q^2}{x}\Big)\right]_{x_{10}^2=\frac{1}{Q^2}}, \\
\label{eq:g1}
g_1\Lb x,Q^2\Rb &= -\sum_f\frac{N_cZ_f^2}{4\pi^3}\int\limits_{\Lambda^2/s}^1\frac{dz}{z}\!\!\int\limits_{\frac{1}{zs}}^{\min\left\{\frac{1}{zQ^2},\frac{1}{\Lambda^2}\right\}}\!\!\frac{dx_{10}^2}{x_{10}^2}\,\big[Q+2G_2\big]\Lb x_{10}^2,zs\Rb,
\end{align}
\end{widetext}
where in Eqs.~(\ref{eq:DSigma}) and (\ref{eq:g1}) the center-of-mass energy is $s=Q^2/x$, $Z_f$ is the electric charge of the quark flavor $f$ in units of the electron charge, and \eq{eq:g1} is the double-logarithmic limit of the full expression~\cite{Cougoulic:2022gbk}. Two features of these formulas are important below. First, $\Delta\Sigma$ and $g_1$ have the same integrand, so at large $N_c$ they have the same $x$ dependence; we discuss only $\Delta\Sigma$, and all statements apply to $g_1$. Second, $\Delta\Sigma$ integrates the amplitudes over all dipole sizes between $1/\sqrt{zs}$ and $\min\{1/\sqrt{zQ^2},1/\Lambda\}$, whereas $\Delta G$ is given by $G_2$ at the single dipole size $x_{10}=1/Q$. Since $x_{10}^2\,\partial/\partial x_{10}^2$ removes one logarithm, at DLA accuracy
\beq
\label{eq:DG_DLA}
\Delta G\Lb x,Q^2\Rb \simeq \frac{2N_c}{\alpha_s\pi^2}\,G_2\Big(x_{10}^2=\frac{1}{Q^2},\,zs=\frac{Q^2}{x}\Big). 
\eeq

\subsection{Large-$N_c$ evolution with the unpolarized $S$-matrices}
\label{sec:largeNc}

\begin{figure}[tbp]
\centering
\begin{tikzpicture}[scale=0.95,>=stealth]
\fill[gray!25] (3.3,-0.1) rectangle (3.75,3.7);
\node[below] at (3.52,-0.1) {\small target};
\draw[thick] (0,3.2) -- (6.2,3.2); \node[left] at (0,3.2) {$\un x_1$};
\draw[thick] (0,0.4) -- (6.2,0.4); \node[left] at (0,0.4) {$\un x_0$};
\draw[thick,decorate,decoration={coil,aspect=0.6,segment length=4pt,amplitude=2.5pt}] (0.9,3.2) -- (2.0,2.2);
\draw[thick] (2.0,2.2) -- (2.15,2.32) -- (6.2,2.32);
\draw[thick] (2.0,2.2) -- (2.15,2.08) -- (6.2,2.08);
\node[right] at (6.2,2.2) {$\un x_2$};
\fill (0.9,3.2) circle (1.6pt);
\draw[fill=white] (3.52,2.32) circle (3.5pt); \draw (3.42,2.22) -- (3.62,2.42); \draw (3.42,2.42) -- (3.62,2.22);
\draw[<->,red!70!black] (5.3,2.37) -- (5.3,3.15); \node[right] at (5.32,2.76) {\small $x_{21}$: polarized};
\draw[<->,blue!70!black] (4.5,0.45) -- (4.5,2.03); \node[right] at (4.52,1.25) {\small $x_{20}$: $S_{20}$};
\draw[<->] (0.45,0.45) -- (0.45,3.15); \node[right] at (0.5,1.8) {\small $x_{10}$};
\end{tikzpicture}
\caption{Schematic picture of one step of the large-$N_c$ helicity evolution in the shock-wave approach. The soft gluon emitted with momentum fraction $z'\ll z$ is a quark--antiquark pair at $\un x_2$ (double line), which splits the parent dipole $10$ into the daughters $21$ and $20$. In the term shown, the helicity-dependent sub-eikonal interaction with the target (cross) belongs to the daughter $21$, while the daughter $20$ interacts eikonally and contributes the unpolarized $S$-matrix $S_{20}$. In the double-logarithmic region $x_{21}\ll x_{10}\simeq x_{20}$ one has $S_{20}\simeq S_{10}$. In the terms with the neighbor amplitudes the roles of the two daughters are exchanged and the unpolarized factor is $S_{21}$, the $S$-matrix of the small daughter.}
\label{fig:diagram}
\end{figure}
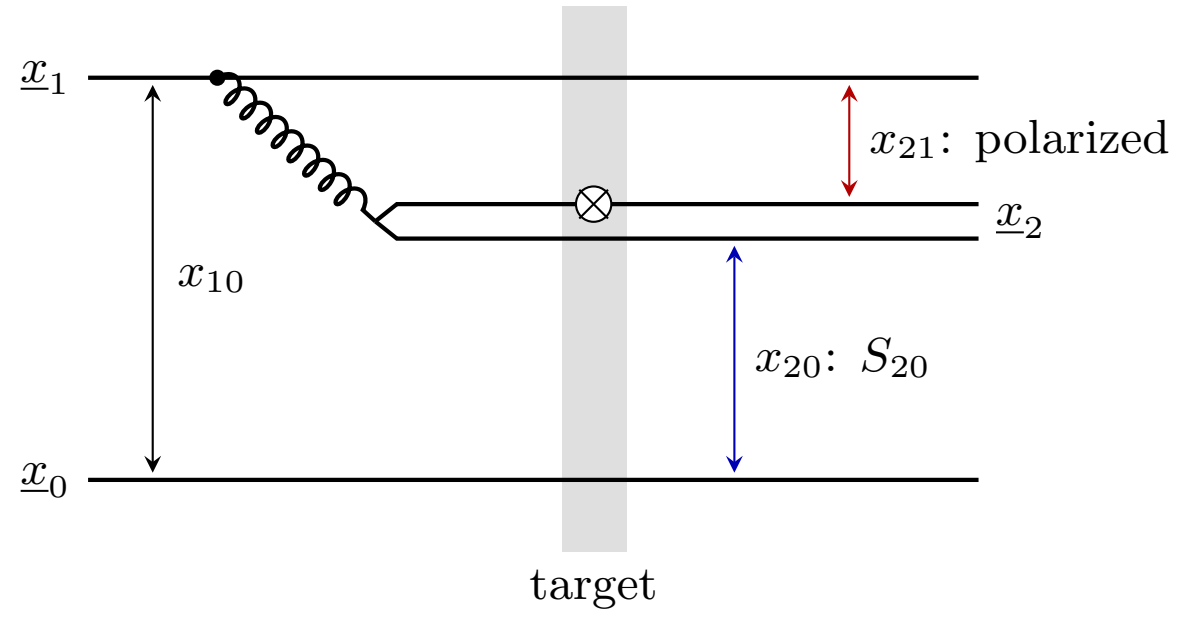

The evolution of the polarized amplitudes is generated by the emission of a soft gluon at the transverse position ${\un x}_2$, with momentum fraction $z'\ll z$, by the projectile dipole $10$. At large $N_c$ the gluon is equivalent to a quark--antiquark pair, and the parent dipole $10$ is replaced by the two daughter dipoles $20$ and $21$ (see Fig.~\ref{fig:diagram}). One of the daughters carries the polarization; the other is unpolarized and interacts with the target eikonally. Its interaction is described by an $S$-matrix~(\ref{eq:Sdef}). This is how the unpolarized $S$-matrices enter the helicity evolution. Explicitly, the large-$N_c$ evolution equation for $G$ reads~\cite{Cougoulic:2022gbk}
\begin{widetext}
\beq
\label{eq:G_largeNc}
\begin{split}
G_{10}\Lb zs\Rb ={}& G_{10}^{\Lb 0\Rb}\Lb zs\Rb + \frac{\alpha_sN_c}{2\pi^2}\int\limits_{\Lambda^2/s}^z\frac{dz'}{z'}\int d^2x_2 \Bigg\{ 2\left[\frac{1}{x_{21}^2}-\frac{{\un x}_{21}\cdot{\un x}_{20}}{x_{21}^2x_{20}^2}\right]\Big[S_{20}\,G_{21}+S_{21}\,\Gamma^{\mathrm{gen}}_{20,21}\Big] \\
&+ \left[2\frac{\epsilon^{ij}x_{21}^j}{x_{21}^4}-\frac{\epsilon^{ij}\Lb x_{20}^j+x_{21}^j\Rb}{x_{20}^2x_{21}^2}-\frac{2\,{\un x}_{20}\times{\un x}_{21}}{x_{20}^2x_{21}^2}\left(\frac{x_{21}^i}{x_{21}^2}-\frac{x_{20}^i}{x_{20}^2}\right)\right]\Big[S_{20}\,G^i_{21}+S_{21}\,\Gamma^{i\,\mathrm{gen}}_{20,21}\Big] \\
&+ \frac{x_{10}^2}{x_{21}^2x_{20}^2}\Big[S_{20}\,G_{12}-\Gamma^{\mathrm{gen}}_{10,21}\Big]\Bigg\},
\end{split}
\eeq
and the one for $G^i$ reads~\cite{Cougoulic:2022gbk}
\beq
\label{eq:Gi_largeNc}
\begin{split}
G^i_{10}\Lb zs\Rb = G^{i\,\Lb 0\Rb}_{10}\Lb zs\Rb &+ \frac{\alpha_sN_c}{2\pi^2}\int\limits_{\Lambda^2/s}^z\frac{dz'}{z'}\int d^2x_2\,\frac{x_{10}^2}{x_{21}^2x_{20}^2}\Big[S_{20}\,G^i_{12}-\Gamma^{i\,\mathrm{gen}}_{10,21}\Big] \\
&+ \frac{\alpha_sN_c}{4\pi^2}\int\limits_{\Lambda^2/s}^z\frac{dz'}{z'}\int d^2x_2\,\Bigg\{2\left[\frac{\epsilon^{ij}x_{21}^j}{x_{21}^2}-\frac{\epsilon^{ij}x_{20}^j}{x_{20}^2}+2x_{21}^i\frac{{\un x}_{21}\times{\un x}_{20}}{x_{21}^2x_{20}^2}\right]\Big[S_{20}\,G_{21}+S_{21}\,\Gamma^{\mathrm{gen}}_{20,21}\Big] \\
&\qquad + K^{ij}\Lb {\un x}_{20},{\un x}_{21}\Rb\,\Big[S_{20}\,G^j_{21}+S_{21}\,\Gamma^{j\,\mathrm{gen}}_{20,21}\Big]\Bigg\},
\end{split}
\eeq
with
\beq
\label{eq:Kij}
\begin{split}
K^{ij}\Lb {\un x}_{20},{\un x}_{21}\Rb = {}& \delta^{ij}\left(\frac{3}{x_{21}^2}-2\frac{{\un x}_{20}\cdot{\un x}_{21}}{x_{20}^2x_{21}^2}-\frac{1}{x_{20}^2}\right) - 2\frac{x_{21}^ix_{20}^j}{x_{21}^2x_{20}^2}\left(2\frac{{\un x}_{20}\cdot{\un x}_{21}}{x_{20}^2}+1\right) \\
&+ 2\frac{x_{21}^ix_{21}^j}{x_{21}^2x_{20}^2}\left(2\frac{{\un x}_{20}\cdot{\un x}_{21}}{x_{21}^2}+1\right) + 2\frac{x_{20}^ix_{20}^j}{x_{20}^4} - 2\frac{x_{21}^ix_{21}^j}{x_{21}^4}.
\end{split}
\eeq
\end{widetext}
Here all the amplitudes and $S$-matrices inside the integrals are evaluated at the rapidity $z's$, and $\Gamma_{20,21}$ and $\Gamma^i_{20,21}$ are the neighbor polarized dipole amplitudes~\cite{Kovchegov:2015pbl,Kovchegov:2016zex,Kovchegov:2017lsr,Kovchegov:2018znm}. Their operator definition is the same as that of $G_{20}$ and $G^i_{20}$, but the subsequent evolution in them is restricted by the lifetime of the adjacent dipole of size $x_{21}$. The generalized amplitudes $\Gamma^{\mathrm{gen}}$ reduce to $\Gamma$ in the region where the neighbor dipole is smaller than the polarized one, and to $G$ otherwise~\cite{Cougoulic:2022gbk}. The lifetime ordering of the successive emissions, $z'x_{21}^2\ll zx_{10}^2$, is implied in all IR-divergent integrals~\cite{Cougoulic:2019aja,Cougoulic:2022gbk}. Finally, the Born-level initial conditions are~\cite{Kovchegov:2016zex,Kovchegov:2017lsr,Cougoulic:2022gbk}
\beq
\label{eq:Born}
\begin{split}
G^{\Lb 0\Rb}\Lb x_{10}^2,zs\Rb &= \frac{\alpha_s^2C_F}{2N_c}\,\pi\left[C_F\ln\frac{zs}{\Lambda^2}-2\ln\Lb zsx_{10}^2\Rb\right], \\
G_2^{\Lb 0\Rb}\Lb x_{10}^2,zs\Rb &= \frac{\alpha_s^2C_F}{N_c}\,\pi\,\ln\frac{1}{x_{10}\Lambda}.
\end{split}
\eeq

Three facts are visible in Eqs.~(\ref{eq:G_largeNc}) and (\ref{eq:Gi_largeNc}). First, every real-emission term carries exactly one $S$-matrix: the polarized daughter is accompanied by the $S$-matrix of the unpolarized one. Second, the virtual terms, $-\Gamma^{\mathrm{gen}}_{10,21}$ and $-\Gamma^{i\,\mathrm{gen}}_{10,21}$, carry no $S$-matrix: in them the parent dipole $10$ is not split. Third, the $S$-matrices are evaluated at the rapidity $z's$ of the emitted gluon. For $S=1$ the real and virtual terms combine in such a way that the ultraviolet divergences at $x_{21}\to0$ cancel. For $S\neq1$ the balance between real and virtual terms is modified. This is the main source of the new terms in the equations of Sec.~\ref{sec:DLA}.

\subsection{Counting of logarithms}
\label{sec:counting}

The unpolarized $S$-matrix evolves with the BK/JIMWLK equations, whose kernel resums powers of $\alpha_s\ln\Lb 1/x\Rb$: the evolution of $S$ is single-logarithmic in the longitudinal momentum (SLA$_L$). In the double-logarithmic approximation, which resums powers of $\alpha_s\ln^2\Lb 1/x\Rb$, one therefore puts $S_{20}=S_{21}=1$ in Eqs.~(\ref{eq:G_largeNc}) and (\ref{eq:Gi_largeNc}). This is the step taken in Ref.~\cite{Cougoulic:2022gbk}, and it leads to the DLA equations solved in Refs.~\cite{Cougoulic:2022gbk,Borden:2023ugd}.

At fixed coupling, however, the DLA parameter $\alpha_s\ln^2\Lb 1/x\Rb$ is of order one already when $\alpha_s\ln\Lb 1/x\Rb\sim\sqrt{\alpha_s}$, which for realistic couplings is not much smaller than one. The asymptotic behavior~(\ref{eq:asymp_intro}) sets in only when $\sqrt{\bas}\ln\Lb 1/x\Rb\gg1$, that is, when the BK evolution parameter $\bas\ln\Lb 1/x\Rb$ is itself large. At such energies the saturation scale has grown appreciably, and the $S$-matrix of a dipole of size $x\gtrsim1/Q_s$ is far from one. Setting $S=1$ is then correct for dipoles much smaller than $1/Q_s$ but not for the larger ones. In this paper we keep the $S$-matrices in the evolution of the polarized amplitudes, treating $S$ as an external function given by the solution of the unpolarized evolution. This combines the DLA evolution of the helicity with the leading-logarithmic evolution of $S$ in a single set of equations. As discussed in Ref.~\cite{Kovchegov:2015pbl}, such a combination is not justified at strict DLA accuracy: only a complete calculation at the next order would decide whether the leading-logarithmic saturation corrections can be kept. It was argued in Ref.~\cite{Kovchegov:2015pbl}, however, that such a calculation is not expected to change the operator structure of the evolution, and that it is therefore likely justified to combine the two, as was done for the Reggeon evolution in Ref.~\cite{Itakura:2003jp}. We adopt this point of view.

We make two further assumptions, both standard in the derivation of the KPS-CTT equations. (i) We work at large $N_c$ and in the DLA for the helicity sector, and we keep only the terms that give double logarithms; the single-logarithmic terms of Ref.~\cite{Kovchegov:2021lvz} are not included. (ii) The amplitudes vary slowly with the impact parameter $b$ on the scale of the dipole sizes involved. This assumption is needed to obtain closed equations for the impact-parameter integrated amplitudes~\cite{Cougoulic:2022gbk}. In our case it implies that the equations hold at each $b$ separately, with $S\Lb x^2,z';b\Rb$ and a local saturation scale $Q_s\Lb b\Rb$. In what follows we suppress the argument $b$; in the numerical part we use a $b$-independent target.

\section{Double-logarithmic evolution with saturation}
\label{sec:DLA}

We now extract the double-logarithmic part of Eqs.~(\ref{eq:G_largeNc}) and (\ref{eq:Gi_largeNc}) while keeping the $S$-matrices. We follow step by step the reduction of Ref.~\cite{Cougoulic:2022gbk} for $S=1$, and point out where the $S$-matrices change it.

A double logarithm requires a logarithmic integral over the transverse size of the daughter dipole in addition to the one over $z'$. In Eqs.~(\ref{eq:G_largeNc}) and (\ref{eq:Gi_largeNc}) the transverse integrals are logarithmic in two regions: the ultraviolet (UV) region $x_{21}\ll x_{10}\simeq x_{20}$, in which the emitted gluon is close to the line $\un x_1$, and the infrared (IR) region $x_{21}\simeq x_{20}\gg x_{10}$, in which both daughters are much larger than the parent. The $S$-matrices in these two regions are
\beq
\label{eq:Sregions}
\begin{split}
&\text{UV:}\quad S_{20}\simeq S\Lb x_{10}^2,z's\Rb,\qquad S_{21}=S\Lb x_{21}^2,z's\Rb, \\
&\text{IR:}\quad S_{20}\simeq S_{21}\simeq S\Lb x_{21}^2,z's\Rb.
\end{split}
\eeq
In the UV region the two $S$-matrices are different: $S_{20}$ is the $S$-matrix of a dipole of the size of the parent, while $S_{21}$ is the $S$-matrix of the small daughter, which is closer to one. Inside the saturation region this difference is large, and it controls the results of Sec.~\ref{sec:asympt}. In the IR region both daughters are large and both $S$-matrices are those of the large daughter.

\subsection{The equation for $G$}
\label{sec:Geq}

In the UV region $x_{21}\ll x_{10}$ the kernels of \eq{eq:G_largeNc} that multiply $G$ and $\Gamma$ both approach $1/x_{21}^2$: the first one is $2/x_{21}^2$, and the BK-type kernel $x_{10}^2/\Lb x_{21}^2x_{20}^2\Rb\to1/x_{21}^2$. In Ref.~\cite{Cougoulic:2022gbk} this region gives the combination $\Gamma+3G$ in the DLA equation for $G$: $2G_{21}+G_{12}\to3G\Lb x_{21}^2\Rb$, using $G_{12}=G_{21}$, and $2\Gamma_{20,21}-\Gamma_{10,21}\to\Gamma\Lb x_{10}^2,x_{21}^2\Rb$, since $x_{20}\simeq x_{10}$. With the $S$-matrices~(\ref{eq:Sregions}) kept, the same terms give
\beq
\label{eq:Gbracket1}
\begin{split}
&2\big[S_{20}G_{21}+S_{21}\Gamma_{20,21}\big] + \big[S_{20}G_{12}-\Gamma_{10,21}\big] \\
&\quad\to 3\,S\Lb x_{10}^2\Rb\,G\Lb x_{21}^2\Rb + \big[2S\Lb x_{21}^2\Rb-1\big]\,\Gamma\Lb x_{10}^2,x_{21}^2\Rb. 
\end{split}
\eeq
The terms with $G^i$ are treated in the same way. After the angular integration and the decomposition~(\ref{eq:decomp}), Ref.~\cite{Cougoulic:2022gbk} obtains $2G_2\Lb x_{21}^2\Rb+2\Gamma_2\Lb x_{10}^2,x_{21}^2\Rb$ from the second line of \eq{eq:G_largeNc}, where the first term comes with $S_{20}$ and the second with $S_{21}$. Hence
\beq
\label{eq:Gbracket2}
\begin{split}
2\big[S_{20}G^i_{21}+S_{21}\Gamma^{i}_{20,21}\big] \to{}& 2\,S\Lb x_{10}^2\Rb\,G_2\Lb x_{21}^2\Rb \\
&+ 2\,S\Lb x_{21}^2\Rb\,\Gamma_2\Lb x_{10}^2,x_{21}^2\Rb. 
\end{split}
\eeq
(We do not write the arguments $z's$.) Adding Eqs.~(\ref{eq:Gbracket1}) and (\ref{eq:Gbracket2}), the bracket of the DLA equation for $G$ becomes
\beq
\label{eq:Gbracket}
\begin{split}
\mathcal{K}_G \equiv{}& \big[2S\Lb x_{21}^2\Rb-1\big]\,\Gamma\Lb x_{10}^2,x_{21}^2\Rb + 3\,S\Lb x_{10}^2\Rb\,G\Lb x_{21}^2\Rb \\
&+ 2\,S\Lb x_{10}^2\Rb\,G_2\Lb x_{21}^2\Rb + 2\,S\Lb x_{21}^2\Rb\,\Gamma_2\Lb x_{10}^2,x_{21}^2\Rb.
\end{split}
\eeq
For $S=1$ this is $\Gamma+3G+2G_2+2\Gamma_2$, the bracket of Ref.~\cite{Cougoulic:2022gbk}. It is convenient to write it as
\beq
\label{eq:Gbracket_alt}
\begin{split}
\mathcal{K}_G ={}& \Gamma + 3SG + 2SG_2 + 2\Gamma_2 \\
&- 2\big[1-S\Lb x_{21}^2\Rb\big]\big(\Gamma+\Gamma_2\big),
\end{split}
\eeq
where now $S$ without argument denotes $S\Lb x_{10}^2\Rb$. In this form the new terms are proportional to $1-S$ and vanish in the dilute regime.

The physical content of \eq{eq:Gbracket} is simple. When the polarization is transferred to the small daughter, the other daughter has the size of the parent, and its $S$-matrix $S\Lb x_{10}^2\Rb$ suppresses the emission if the parent is saturated. When the polarization stays in a dipole of the size of the parent (the neighbor amplitudes), the unpolarized daughter is the small one, and its $S$-matrix $S\Lb x_{21}^2\Rb$ is close to one unless the small daughter is also saturated. The virtual term $-\Gamma$ carries no $S$-matrix. If both the parent and the daughter are saturated, it is the only term that survives in the coefficient of $\Gamma$, which then changes sign.

In the IR region the kernels of the $G$ equation fall off as $x_{10}/x_{21}^3$ or faster, and they do not produce a logarithm, with or without $S$. The remaining region in which a logarithm could arise is $x_{20}\ll x_{10}\simeq x_{21}$. There the BK-type kernel behaves as $1/x_{20}^2$ and multiplies $S_{20}G_{12}-\Gamma^{\mathrm{gen}}_{10,21}$. For $x_{20}\to0$ one has $x_{21}\to x_{10}$, so $G_{12}\to G_{10}$ and $\Gamma^{\mathrm{gen}}_{10,21}\to G_{10}$, while the dipole $20$ is small and $S_{20}\to1$. The bracket vanishes, and no logarithm is generated, exactly as for $S=1$. We conclude that saturation does not create new double-logarithmic regions in the $G$ equation.

\subsection{The equation for $G_2$}
\label{sec:G2eq}

The DLA equation for $G_2$ is obtained from \eq{eq:Gi_largeNc} after contracting with $\epsilon^{ij}x_{10}^j$ and integrating over the impact parameter. For $S=1$ the result is the DLA equation for $G_2$ of Ref.~\cite{Cougoulic:2022gbk}. It has contributions from two regions.

\emph{IR region.} For $x_{21}\simeq x_{20}\gg x_{10}$ the terms in the second integral of \eq{eq:Gi_largeNc} give a logarithmic integral over the size of the large daughters, with the combination $2[G+2G_2]\Lb x_{21}^2\Rb$~\cite{Cougoulic:2022gbk}. According to \eq{eq:Sregions}, all of them now carry the factor $S\Lb x_{21}^2,z's\Rb$:
\beq
\label{eq:G2IR}
2\big[G+2G_2\big]\Lb x_{21}^2,z's\Rb \to 2\,S\Lb x_{21}^2,z's\Rb\big[G+2G_2\big]\Lb x_{21}^2,z's\Rb. 
\eeq
For $S=1$ the size of the large daughters is cut off by the infrared cutoff $1/\Lambda$. With saturation it is effectively cut off at $x_{21}\sim1/Q_s\Lb z's\Rb$, beyond which $S$ vanishes.

\emph{UV region.} For $x_{21}\to0$ two terms of \eq{eq:Gi_largeNc} behave as $1/x_{21}^2$ when multiplied by the neighbor amplitude $\Gamma_2\Lb x_{10}^2,x_{21}^2\Rb$: the real term with $K^{ij}$, which carries $S_{21}$, and the virtual term $-\Gamma^{i\,\mathrm{gen}}_{10,21}$ in the first line, which carries no $S$-matrix. In Ref.~\cite{Cougoulic:2022gbk} it is stated that the two UV divergences cancel. We have verified this cancellation explicitly. After contracting with $\epsilon^{ij}x_{10}^j/x_{10}^2$ and averaging over the direction of ${\un x}_{21}$, the two kernels behave, in units of $\Lb \alpha_sN_c/2\pi^2\Rb\,\Lb 1/x_{21}^2\Rb$, as
\beq
\label{eq:UVcoef}
\text{real: } +1\times S\Lb x_{21}^2\Rb, \qquad \text{virtual: } -1,
\eeq
up to corrections suppressed by powers of $x_{21}/x_{10}$ (Appendix~\ref{app:UV}). For $S=1$ they cancel. For $S\neq1$ they leave
\beq
\label{eq:G2UV}
\begin{split}
-\frac{\alpha_sN_c}{2\pi^2}\int\frac{dz'}{z'}\int\limits_{x_{21}<x_{10}}\frac{d^2x_{21}}{x_{21}^2}\,&\big[1-S\Lb x_{21}^2,z's\Rb\big] \\
&\times\Gamma_2\Lb x_{10}^2,x_{21}^2,z's\Rb.
\end{split}
\eeq
This term is new. It vanishes when the small daughters are not saturated, and it is negative when they are. Since the two original terms were never needed separately for $S=1$, their integration limits were not written in Ref.~\cite{Cougoulic:2022gbk}. They follow from the kinematics of the UV region, which is the same as in the $G$ equation: the daughter is smaller than the parent, $x_{21}<x_{10}$, and the lifetime ordering $1/s\ll z'x_{21}^2\ll zx_{10}^2$ holds. The second inequality is automatic for $z'<z$ and $x_{21}<x_{10}$. In logarithmic variables these conditions give the same limits as the UV integral of the $G$ equation, see \eq{eq:system} below.\\

\subsection{The final system}
\label{sec:system}

We use the variables of Ref.~\cite{Cougoulic:2022gbk},
\beq
\label{eq:vars}
\eta = \sqrt{\bas}\,\ln\frac{zs}{\Lambda^2}, \qquad s_{10} = \sqrt{\bas}\,\ln\frac{1}{x_{10}^2\Lambda^2},
\eeq
and similarly $\eta'$, $\eta''$, $s_{21}$, $s_{32}$. The unpolarized $S$-matrix becomes a function $S\Lb s,\eta\Rb$ of the logarithmic size and rapidity of the dipole. Collecting Eqs.~(\ref{eq:Gbracket}), (\ref{eq:G2IR}) and (\ref{eq:G2UV}), and the analogous equations for the neighbor amplitudes, we obtain
\begin{widetext}
\begin{subequations}
\label{eq:system}
\begin{align}
G\Lb s_{10},\eta\Rb ={}& G^{\Lb 0\Rb}\Lb s_{10},\eta\Rb + \int\limits_{s_{10}}^{\eta}d\eta'\int\limits_{s_{10}}^{\eta'}ds_{21}\,\Big\{\big[2S\Lb s_{21},\eta'\Rb-1\big]\Gamma\Lb s_{10},s_{21},\eta'\Rb + 3S\Lb s_{10},\eta'\Rb\,G\Lb s_{21},\eta'\Rb \notag \\
&\hspace{4.6cm} + 2S\Lb s_{10},\eta'\Rb\,G_2\Lb s_{21},\eta'\Rb + 2S\Lb s_{21},\eta'\Rb\,\Gamma_2\Lb s_{10},s_{21},\eta'\Rb\Big\}, \label{eq:sysG} \\
\Gamma\Lb s_{10},s_{21},\eta'\Rb ={}& G^{\Lb 0\Rb}\Lb s_{10},\eta'\Rb + \int\limits_{s_{10}}^{\eta'}d\eta''\!\!\!\int\limits_{\max[s_{10},\,s_{21}+\eta''-\eta']}^{\eta''}\!\!\!ds_{32}\,\Big\{\big[2S\Lb s_{32},\eta''\Rb-1\big]\Gamma\Lb s_{10},s_{32},\eta''\Rb + 3S\Lb s_{10},\eta''\Rb\,G\Lb s_{32},\eta''\Rb \notag \\
&\hspace{4.6cm} + 2S\Lb s_{10},\eta''\Rb\,G_2\Lb s_{32},\eta''\Rb + 2S\Lb s_{32},\eta''\Rb\,\Gamma_2\Lb s_{10},s_{32},\eta''\Rb\Big\}, \label{eq:sysGam} \\
G_2\Lb s_{10},\eta\Rb ={}& G_2^{\Lb 0\Rb}\Lb s_{10},\eta\Rb + 2\int\limits_0^{s_{10}}ds_{21}\!\!\int\limits_{s_{21}}^{\eta-s_{10}+s_{21}}\!\!d\eta'\,S\Lb s_{21},\eta'\Rb\,\big[G\Lb s_{21},\eta'\Rb+2G_2\Lb s_{21},\eta'\Rb\big] \notag \\
& - \int\limits_{s_{10}}^{\eta}d\eta'\int\limits_{s_{10}}^{\eta'}ds_{21}\,\big[1-S\Lb s_{21},\eta'\Rb\big]\,\Gamma_2\Lb s_{10},s_{21},\eta'\Rb, \label{eq:sysG2} \\
\Gamma_2\Lb s_{10},s_{21},\eta'\Rb ={}& G_2^{\Lb 0\Rb}\Lb s_{10},\eta'\Rb + 2\int\limits_0^{s_{10}}ds_{32}\!\!\int\limits_{s_{32}}^{\eta'-s_{21}+s_{32}}\!\!d\eta''\,S\Lb s_{32},\eta''\Rb\,\big[G\Lb s_{32},\eta''\Rb+2G_2\Lb s_{32},\eta''\Rb\big] \notag \\
& - \int\limits_{s_{10}}^{\eta'}d\eta''\!\!\!\int\limits_{\max[s_{10},\,s_{21}+\eta''-\eta']}^{\eta''}\!\!\!ds_{32}\,\big[1-S\Lb s_{32},\eta''\Rb\big]\,\Gamma_2\Lb s_{10},s_{32},\eta''\Rb. \label{eq:sysGam2}
\end{align}
\end{subequations}
\end{widetext}
The neighbor amplitudes are defined for $s_{21}\ge s_{10}$ (the neighbor dipole is smaller than the polarized one), and on the diagonal $\Gamma\Lb s_{10},s_{10},\eta\Rb=G\Lb s_{10},\eta\Rb$, $\Gamma_2\Lb s_{10},s_{10},\eta\Rb=G_2\Lb s_{10},\eta\Rb$. The limits of the integrals in the neighbor equations implement the lifetime ordering with respect to the neighbor dipole, exactly as in Ref.~\cite{Cougoulic:2022gbk}. In the UV term of \eq{eq:sysGam2} the limits follow by the same argument as in Sec.~\ref{sec:G2eq}: the evolution of the neighbor amplitude is bounded by the lifetime $z'x_{21}^2$ of the neighbor dipole, which gives the same limits as in \eq{eq:sysGam}. The inhomogeneous terms are the Born terms~(\ref{eq:Born}), which in these variables read~\cite{Cougoulic:2022gbk}
\beq
\label{eq:Born_vars}
\begin{split}
G^{\Lb 0\Rb}\Lb s_{10},\eta\Rb &= \mathcal{N}\,\big[C_F\,\eta-2\Lb \eta-s_{10}\Rb\big], \\
G_2^{\Lb 0\Rb}\Lb s_{10},\eta\Rb &= \mathcal{N}\,s_{10},
\end{split}
\eeq
with $\mathcal{N}=\pi\alpha_s^2C_F/\Lb 2N_c\sqrt{\bas}\Rb$. Since the equations are linear in the polarized amplitudes, the normalization $\mathcal{N}$ drops out of all the ratios and exponents discussed below. Note that the Born terms do not contain $S$; a saturation-modified initial condition, for instance one obtained in the quasi-classical approach of Ref.~\cite{Li:2024xra}, would change the magnitudes at moderate $x$ but not the asymptotic exponents, which are properties of the kernel.

The system~(\ref{eq:system}) is our first main result. Its properties are:
\begin{enumerate}
\item For $S=1$ it reduces exactly to the DLA equations of Ref.~\cite{Cougoulic:2022gbk} written in the variables~(\ref{eq:vars}). We have also checked that our numerical solution of \eq{eq:system} with $S=1$ coincides, within round-off, with an independent solution of the equations of Ref.~\cite{Cougoulic:2022gbk} (Sec.~\ref{sec:validation} and Appendix~\ref{app:numerics}).
\item Saturation enters in two ways: through the factors $S$ that multiply the real emissions, and through new terms proportional to $1-S$, namely $-2[1-S\Lb s_{21}\Rb]\Lb \Gamma+\Gamma_2\Rb$ in the equation for $G$ and $-[1-S\Lb s_{21}\Rb]\Gamma_2$ in the equation for $G_2$. The latter is the remnant of the incomplete cancellation of the UV divergences.
\item The $S$-matrix of the parent, $S\Lb s_{10},\eta'\Rb$, multiplies the emissions in which the polarized daughter is small. The $S$-matrix of the small daughter, $S\Lb s_{21},\eta'\Rb$, multiplies the emissions in which the polarized dipole has the size of the parent, and the new terms. The $S$-matrix of the large daughters, $S\Lb s_{21},\eta'\Rb$ with $s_{21}<s_{10}$, multiplies the IR term of the $G_2$ equation.
\item The single-logarithmic terms are not included, as in Ref.~\cite{Cougoulic:2022gbk}. The derivation in Secs.~\ref{sec:Geq} and \ref{sec:G2eq} shows that saturation does not generate new double-logarithmic regions.
\end{enumerate}

\subsection{The saturation boundary}
\label{sec:boundary}

\begin{figure}[tbp]
\centering
\includegraphics[width=\columnwidth]{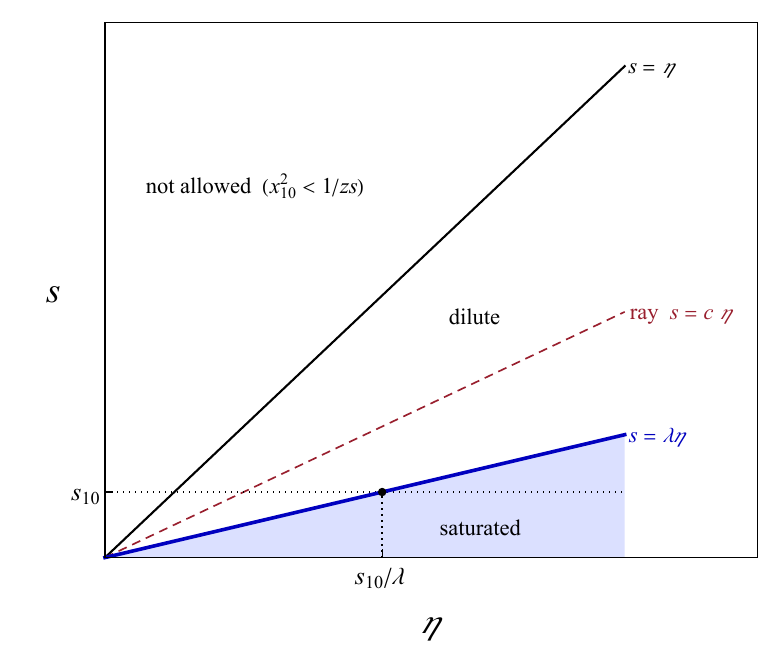}
\caption{The $\Lb s,\eta\Rb$ plane. The polarized amplitudes are defined for $0\le s\le\eta$. Dipoles below the line $s=\lambda\eta$ are saturated. A dipole of fixed size $s_{10}$ enters the saturation region at $\eta=s_{10}/\lambda$. The growth rate of the amplitudes along the rays $s=c\,\eta$ is discussed in Sec.~\ref{sec:edge}.}
\label{fig:plane}
\end{figure}

The $S$-matrix enters \eq{eq:system} as an external function. Its main feature is the saturation scale. For $Q_s^2\Lb zs\Rb=\Lambda^2\,\Lb zs/\Lambda^2\Rb^{\lambda}$ a dipole is saturated when $x^2Q_s^2\gtrsim1$, that is, in the variables~(\ref{eq:vars}), when
\beq
\label{eq:satline}
s < s_{\mathrm{sat}}\Lb \eta\Rb = \lambda\,\eta .
\eeq
The saturation region is the wedge below the straight line $s=\lambda\eta$ in the $\Lb s,\eta\Rb$ plane (Fig.~\ref{fig:plane}). Note that $s$ and $\eta$ are both measured in units of $1/\sqrt{\bas}$, so the slope of the line is the usual exponent $\lambda$ of the saturation scale. For example, in the GBW model~\cite{Golec-Biernat:1998zce}
\beq
\label{eq:GBW}
S\Lb s,\eta\Rb = \exp\Big[-\tfrac14\,x^2Q_s^2\Big] = \exp\Big[-\tfrac14\,e^{\Lb \lambda\eta-s\Rb/\sqrt{\bas}}\Big],
\eeq
which is a step of width of order $\sqrt{\bas}$ around the line $s=\lambda\eta$. In the strict DLA limit $\sqrt{\bas}\to0$ at fixed $s$ and $\eta$, it becomes a sharp boundary,
\beq
\label{eq:step}
S\Lb s,\eta\Rb \to \theta\Lb s-\lambda\eta\Rb.
\eeq
The same is true for the solution of the BK equation, which has the same geometric-scaling form~\cite{Stasto:2000er,Iancu:2002tr,Mueller:2002zm}, with a smooth but narrow transition region. The polarized amplitudes grow along the $\eta$ direction with a rate of order $\alpha_h\approx3.66$, while the boundary has a much smaller slope, $\lambda\approx0.2$--$0.3$ in phenomenology~\cite{Golec-Biernat:1998zce,Albacete:2009fh}. A dipole of fixed size $s_{10}$ is dilute for $\eta<s_{10}/\lambda$ and saturated for $\eta>s_{10}/\lambda$.

\section{Asymptotic regime}
\label{sec:asympt}

In this section we study the asymptotic behavior of the solution of \eq{eq:system}, first for a dipole of fixed size $s_{10}$ deep inside the saturation region, $\eta\gg s_{10}/\lambda$, and then for the helicity distributions. We use a saturation boundary with a constant slope $\lambda$, either sharp, \eq{eq:step}, or of the GBW form, \eq{eq:GBW}. The results of this section concern the formal limit of large $\eta$, beyond the values that correspond to the range of $x$ reached by experiments; they characterize the mathematical structure of the evolution. The behavior in the experimental range of $x$ is studied numerically in Sec.~\ref{sec:numerics}.

\subsection{The reduced system}
\label{sec:reduced}

When the parent dipole is saturated, $S\Lb s_{10},\eta'\Rb=0$, and the terms $3S\Lb s_{10}\Rb G$ and $2S\Lb s_{10}\Rb G_2$ disappear from Eqs.~(\ref{eq:sysG}) and (\ref{eq:sysGam}). In Eqs.~(\ref{eq:sysG2}) and (\ref{eq:sysGam2}) all the large daughters, $s_{21}<s_{10}$, are saturated too, and the IR terms vanish. As we show in Sec.~\ref{sec:G2inside}, $G_2$ and $\Gamma_2$ then stop growing, and to exponential accuracy the term $2S\Lb s_{21}\Rb\Gamma_2$ can be dropped from the equation for $\Gamma$. What remains is a closed equation for the neighbor amplitude $\Gamma$,
\beq
\label{eq:redGam}
\begin{split}
\Gamma\Lb s_{10},s_{21},\eta'\Rb \simeq{}& G^{\Lb 0\Rb} + \int\limits_{s_{10}}^{\eta'}d\eta''\!\!\!\int\limits_{\max[s_{10},\,s_{21}+\eta''-\eta']}^{\eta''}\!\!\!ds_{32} \\
&\times\sigma\Lb s_{32},\eta''\Rb\,\Gamma\Lb s_{10},s_{32},\eta''\Rb,
\end{split}
\eeq
with
\beq
\label{eq:sigma}
\sigma\Lb s,\eta\Rb = 2S\Lb s,\eta\Rb-1 = \begin{cases} +1, & s>\lambda\eta\ \text{(dilute)}, \\ -1, & s<\lambda\eta\ \text{(saturated)}. \end{cases}
\eeq
The amplitude $G$ is the value of $\Gamma$ on the diagonal, $G\Lb s_{10},\eta\Rb=\Gamma\Lb s_{10},s_{10},\eta\Rb$. Equation~(\ref{eq:redGam}) has a transparent meaning: a dilute small daughter feeds the growth of the polarized parent, while a saturated one removes it through the virtual term.

It is convenient to introduce the variables
\beq
\label{eq:uv}
u = \eta''-s_{10}, \qquad v = \eta''-s_{32},
\eeq
so that $u$ is the rapidity interval available since the parent was produced, and $v$ is the lifetime variable of the daughter. Writing $\Gamma\Lb s_{10},s_{32},\eta''\Rb\equiv\tilde\Gamma\Lb u,v\Rb$, the integration region of \eq{eq:redGam} becomes $0\le u\le U$, $0\le v\le\min\Lb u,V\Rb$, with $U=\eta'-s_{10}$ and $V=\eta'-s_{21}$, and
\beq
\label{eq:redGam_uv}
\tilde\Gamma\Lb U,V\Rb = G^{\Lb 0\Rb} + \int\limits_0^U du\int\limits_0^{\min\Lb u,V\Rb}dv\,\sigma\Lb u,v\Rb\,\tilde\Gamma\Lb u,v\Rb.
\eeq
For $V<U$ this is equivalent to the hyperbolic equation
\beq
\label{eq:hyperbolic}
\frac{\partial^2\tilde\Gamma}{\partial U\,\partial V} = \sigma\Lb U,V\Rb\,\tilde\Gamma, 
\eeq
with $G=\tilde\Gamma\Lb U,U\Rb$. The integral form~(\ref{eq:redGam_uv}) also fixes the boundary conditions. On the characteristic $V=0$ one has $\tilde\Gamma\Lb U,0\Rb=G^{\Lb 0\Rb}$, and differentiating \eq{eq:redGam_uv} with respect to $V$ gives $\partial_V\tilde\Gamma\Lb U,V\Rb=\int_V^U du\,\sigma\Lb u,V\Rb\,\tilde\Gamma\Lb u,V\Rb$, which vanishes on the diagonal:
\beq
\label{eq:bc}
\tilde\Gamma\Lb U,0\Rb=G^{\Lb 0\Rb}, \qquad \partial_V\tilde\Gamma\Lb U,V\Rb\big|_{V=U}=0 . 
\eeq
The second condition, which is not a Goursat condition, is the trace of the neighbor structure of the KPS-CTT equations: the evolution of the neighbor amplitude is cut off by the lifetime of the neighbor dipole. The daughter is saturated when $s_{32}<\lambda\eta''$, i.e., for $v>\Lb 1-\lambda\Rb\eta''$. For $u\gg s_{10}$ this is
\beq
\label{eq:sigma_uv}
\sigma\Lb u,v\Rb = \begin{cases} +1, & v<\Lb 1-\lambda\Rb\,u, \\ -1, & v>\Lb 1-\lambda\Rb\,u. \end{cases}
\eeq
In the $\Lb u,v\Rb$ plane the region $0\le v\le u$ is divided by the line $v=\Lb 1-\lambda\Rb u$ into a growing wedge ($\sigma=+1$) and an oscillating wedge ($\sigma=-1$). The amplitude $G$ itself lives on the edge $v=u$, which belongs to the oscillating wedge for any $\lambda>0$.

\subsection{Exact solutions for $\lambda\to0^+$ and $\lambda=1$}
\label{sec:exact}

The two limiting cases of \eq{eq:sigma_uv} can be solved exactly. We take a constant inhomogeneous term, normalized to one; the Born term~(\ref{eq:Born_vars}) grows only linearly and does not change the exponents.

\emph{The functions $f_n$.} For any integer $n$, the functions
\beq
\label{eq:fn}
\begin{split}
f_n\Lb U,V\Rb &= \Big(\frac{V}{U}\Big)^{n/2} I_n\big(2\sqrt{UV}\big), \\
g_n\Lb U,V\Rb &= \Big(\frac{V}{U}\Big)^{n/2} J_n\big(2\sqrt{UV}\big)
\end{split}
\eeq
solve $\partial_U\partial_Vf_n=f_n$ and $\partial_U\partial_Vg_n=-g_n$, respectively. They obey the ladder relations
\beq
\label{eq:ladder}
\partial_Vf_n = f_{n-1}, \qquad \partial_Vg_n = g_{n-1},
\eeq
with $f_{-1}=\Lb U/V\Rb^{1/2}I_1$ and $g_{-1}=-\Lb U/V\Rb^{1/2}J_1$, which follow from $d[x^{n}I_n\Lb x\Rb]/dx=x^nI_{n-1}\Lb x\Rb$ and the analogous relation for $J_n$. For $n\ge1$ they vanish at $V=0$, and on the diagonal they reduce to $f_n\Lb U,U\Rb=I_n\Lb 2U\Rb$, $g_n\Lb U,U\Rb=J_n\Lb 2U\Rb$.

\emph{$\lambda\to0^+$.} Consider first the situation in which the parent is saturated but all its small daughters are dilute, so that $\sigma=+1$ everywhere. We look for the solution as $\tilde\Gamma=f_0+\sum_{n\ge1}a_nf_n$, which satisfies the equation and the first condition~(\ref{eq:bc}) for any $a_n$. By \eq{eq:ladder} the second condition reads $I_1\Lb 2U\Rb+\sum_{n\ge1}a_nI_{n-1}\Lb 2U\Rb=0$ for all $U$, which is solved by $a_2=-1$ and all other $a_n=0$. Hence
\beq
\label{eq:I0}
\tilde\Gamma\Lb U,V\Rb = I_0\big(2\sqrt{UV}\big) - \frac{V}{U}\,I_2\big(2\sqrt{UV}\big),
\eeq
and, using $I_0\Lb x\Rb-I_2\Lb x\Rb=2I_1\Lb x\Rb/x$,
\beq
\label{eq:Gexact0}
G = \tilde\Gamma\Lb U,U\Rb = \frac{I_1\Lb 2U\Rb}{U} \simeq \frac{e^{2U}}{2\sqrt{\pi}\,U^{3/2}}. 
\eeq
Hence, in this limit,
\beq
\label{eq:omega0}
G\Lb s_{10},\eta\Rb \sim e^{2\eta}\sim\left(\frac{zs}{\Lambda^2}\right)^{2\sqrt{\bas}},
\eeq
with the intercept exactly $2\sqrt{\bas}$ instead of $3.66074\sqrt{\bas}$. We have checked Eqs.~(\ref{eq:I0}) and (\ref{eq:Gexact0}) against the numerical solution of \eq{eq:redGam} with $\sigma=1$: after extrapolation to zero step, $G$ agrees with $I_1\Lb 2U\Rb/U$ to better than $6\times10^{-4}$ (relative) for $U=2$--$10$, and $\tilde\Gamma$ agrees with \eq{eq:I0} at interior points to $5\times10^{-4}$. The growth rate extracted from the numerical solution is $1.9964$, $1.9989$ and $1.9995$ for $\delta=0.1$, $0.05$ and $0.025$, converging to $2$. Note that the quarter-plane solution $I_0\Lb 2\sqrt{UV}\Rb$ has the same exponent but violates the diagonal condition~(\ref{eq:bc}); the neighbor structure changes the power of $U$ in the prefactor from $U^{-1/2}$ to $U^{-3/2}$, not the exponent.

The reduction of the intercept from $3.66$ to $2$ has a simple origin: the terms $3G$ and $2G_2$, which in the KPS-CTT equations describe the transfer of polarization to the small daughter, are switched off by the $S$-matrix of the saturated parent, and only the neighbor amplitude $\Gamma$ survives.

\emph{$\lambda=1$.} In the opposite limit all the small daughters are saturated and $\sigma=-1$ everywhere. The same construction with $g_n$ gives $a_2=+1$,
\beq
\label{eq:Gexact1}
\tilde\Gamma\Lb U,V\Rb = J_0\big(2\sqrt{UV}\big) + \frac{V}{U}\,J_2\big(2\sqrt{UV}\big), \qquad G = \frac{J_1\Lb 2U\Rb}{U}, 
\eeq
which oscillates and decreases as $U^{-3/2}$. The intercept vanishes.

These two exact results fix the end points of the intercept as a function of $\lambda$: $\omega\Lb 0^+\Rb=2$ and $\omega\Lb 1\Rb=0$.

\subsection{$\lambda>0$: the WKB solution}
\label{sec:WKB}

For $\lambda>0$ we look for the asymptotic solution in the form
\beq
\label{eq:WKBansatz}
\tilde\Gamma\Lb U,V\Rb \sim \exp\big[U\,\phi\Lb \rho\Rb\big], \qquad \rho = \frac{V}{U}.
\eeq
At leading order in $1/U$, $\partial_U\partial_V\tilde\Gamma = \phi'\Lb \phi-\rho\phi'\Rb\,\tilde\Gamma$, and \eq{eq:hyperbolic} gives
\beq
\label{eq:eikonal}
\phi'\Lb \rho\Rb\,\big[\phi\Lb \rho\Rb-\rho\,\phi'\Lb \rho\Rb\big] = \sigma .
\eeq
In the growing wedge, $\rho<1-\lambda$, the solution with $\phi\Lb 0\Rb=0$ is
\beq
\label{eq:phi_growing}
\phi\Lb \rho\Rb = 2\sqrt{\rho}, 
\eeq
which is the asymptotic form of the Bessel functions~(\ref{eq:fn}), $f_n\sim e^{2\sqrt{UV}}$. In the oscillating wedge, $\sigma=-1$, the eikonal equation $S_US_V=-1$ for $S=U\phi$ requires $S_U$ and $S_V$ of opposite signs: a solution that grows in $U$ decreases in $V$, as $e^{aU-V/a}$. Such a solution cannot satisfy the diagonal condition~(\ref{eq:bc}), which at this order requires $S_V=0$. The solution in the oscillating wedge is therefore a superposition of oscillating modes, as in the exact case~(\ref{eq:Gexact1}), and it does not generate exponential growth of its own. It transports, with oscillations, what enters it from the boundary $\rho=1-\lambda$ of the growing wedge. The growth rate of $G=\tilde\Gamma\Lb U,U\Rb$ is then set by the largest exponent available in the growing wedge, which is reached at its edge:
\beq
\label{eq:omega_lambda}
\omega\Lb \lambda\Rb = \phi\Lb 1-\lambda\Rb = 2\sqrt{1-\lambda}, \qquad G\Lb s_{10},\eta\Rb \sim e^{\omega\Lb \lambda\Rb\,\eta}. 
\eeq
In physical units the intercept of $G$ inside the saturation region is
\beq
\label{eq:omega_phys}
\alpha_h^{\mathrm{sat}} = 2\sqrt{1-\lambda}\,\sqrt{\bas},
\eeq
which interpolates between $2\sqrt{\bas}$ for $\lambda\to0^+$ and zero for $\lambda\to1$.

Equation~(\ref{eq:omega_lambda}) reproduces the two exact end points, $\omega\Lb 0^+\Rb=2$ [\eq{eq:Gexact0}] and $\omega\Lb 1\Rb=0$ [\eq{eq:Gexact1}]. For $0<\lambda<1$ the boundary between the two wedges is the ray $V=\Lb 1-\lambda\Rb U$, which is not a characteristic, and the problem does not separate in the variables $2\sqrt{UV}$ and $V/U$; we have not found a closed-form solution in this case. Equation~(\ref{eq:omega_lambda}) is therefore a WKB result between two exact end points, not a rigorous derivation, and we have checked it numerically, both for the reduced equation~(\ref{eq:redGam}) and for the full system~(\ref{eq:system}) with the step~(\ref{eq:step}) at $s_{10}=1$. The growth rates, measured as the slope of the envelope of $\ln|G|$, are listed in Table~\ref{tab:omega} and shown in Fig.~\ref{fig:inside}(a). For the reduced system at $\eta=200$--$300$ the agreement with $2\sqrt{1-\lambda}$ is better than $1\%$ for $\lambda\ge0.1$. At small $\lambda$ the convergence is slow, because a dipole of size $s_{10}$ saturates only at $\eta=s_{10}/\lambda$. For example, for $\lambda=0.1$ the measured rate is $1.835$ for $\eta=50$--$100$ and $1.889$--$1.898$ for $\eta=200$--$300$, depending on the step, to be compared with $1.897$. The full system follows the same formula within the accuracy allowed by its smaller range of $\eta$. This confirms that the terms dropped in \eq{eq:redGam} do not change the exponent.

\begin{table}[tbp]
\caption{Growth rate of $G$ inside the saturation region, in units of $\sqrt{\bas}$, compared with \eq{eq:omega_lambda}. Reduced system: \eq{eq:redGam}, $\eta=200$--$300$. Full system: \eq{eq:system} with $S=\theta\Lb s-\lambda\eta\Rb$, $s_{10}=1$, $\delta=0.1$, $\eta=50$--$70$.}
\label{tab:omega}
\begin{ruledtabular}
\begin{tabular}{lccccccc}
$\lambda$ & 0.05 & 0.1 & 0.25 & 0.4 & 0.5 & 0.6 & 0.75 \\ \hline
$2\sqrt{1-\lambda}$ & 1.949 & 1.897 & 1.732 & 1.549 & 1.414 & 1.265 & 1.000 \\
Reduced & 1.927 & 1.889 & 1.729 & 1.550 & 1.418 & 1.273 & 1.007 \\
Full & 1.90 & 1.83 & 1.74 & 1.56 & 1.44 & 1.29 & 1.06
\end{tabular}
\end{ruledtabular}
\end{table}

\subsection{$G_2$ inside the saturation region}
\label{sec:G2inside}

Inside the saturation region the source of $G_2$ disappears. In \eq{eq:sysG2} the IR term is fed by the daughters larger than the parent; if the parent is saturated, so are they, and $S\Lb s_{21},\eta'\Rb=0$. Only the new UV term survives, and the same holds for $\Gamma_2$ in \eq{eq:sysGam2}. In the variables~(\ref{eq:uv}) this gives
\beq
\label{eq:G2red}
\frac{\partial^2\tilde\Gamma_2}{\partial U\,\partial V} = -\,\theta\big(V-\Lb 1-\lambda\Rb U\big)\,\tilde\Gamma_2 ,
\eeq
with $G_2=\tilde\Gamma_2\Lb U,U\Rb$. with $G_2=\tilde\Gamma_2\Lb U,U\Rb$ and the same boundary conditions~(\ref{eq:bc}), since the UV terms of Eqs.~(\ref{eq:sysG2}) and (\ref{eq:sysGam2}) have the same limits as those of the $\Gamma$ equation. The right-hand side is never positive, so \eq{eq:G2red} has no exponentially growing solutions. In the limit $\lambda\to1$, in which all small daughters are saturated, it coincides with the case $\sigma=-1$ of Sec.~\ref{sec:exact}, and
\beq
\label{eq:J0}
G_2 = \frac{J_1\Lb 2U\Rb}{U}, 
\eeq
which oscillates with an amplitude that decreases as $U^{-3/2}$. For $\lambda<1$ the term acts only in the wedge $V>\Lb 1-\lambda\Rb U$, and the decrease is slower but the oscillations are still present. In Figs.~\ref{fig:fixeddipole}(b), \ref{fig:fixeddipole}(d) and \ref{fig:inside}(b) we plot $G_2$ itself, \emph{with its sign}, and not $|G_2|$ or $\ln|G_2|$ as is usual for quantities that grow exponentially. The reason is that $G_2$ changes sign inside the saturation region: on a logarithmic scale each zero of $G_2$ would appear as a narrow downward spike of $\ln|G_2|$, and the oscillation around zero, which is the physical content of Eqs.~(\ref{eq:G2red}) and (\ref{eq:J0}), would be hidden. On a linear scale with the sign kept, the decrease of $G_2$ and its oscillations around zero are directly visible. Figure~\ref{fig:inside}(b) shows the numerical solution of \eq{eq:G2red} for $\lambda=1$, $0.5$ and $0.25$. For $\lambda=1$ it coincides with $J_1\Lb 2U\Rb/U$ to within $5\times10^{-6}$ after extrapolation to zero step. We conclude that $G_2$, and with it $\Delta G$ through \eq{eq:DG_DLA}, stops growing once the dipole of size $1/Q$ is saturated, and then oscillates around zero with a decreasing amplitude. This is the double-logarithmic analogue of the Levin--Tuchin regime~\cite{Levin:1999mw,Levin:2001cv} of the unpolarized amplitude. At single-logarithmic accuracy a monotonic decrease would be expected instead of oscillations; Li found such a Levin--Tuchin behavior for the sub-eikonal dipole gluon distribution at single-logarithmic accuracy~\cite{Li:2026azt}. In the full system the slope of $\ln|G_2|$ at $s_{10}=1$ and $\eta=50$--$70$ is between $-0.04$ and $-0.02$ for $\lambda=0.25$--$0.75$, i.e., $G_2$ does not grow, in agreement with this analysis.

\subsection{Change of regime and the limit $\lambda\to0$}
\label{sec:regime}

\begin{figure*}[tbp]
\centering
\includegraphics[width=\textwidth]{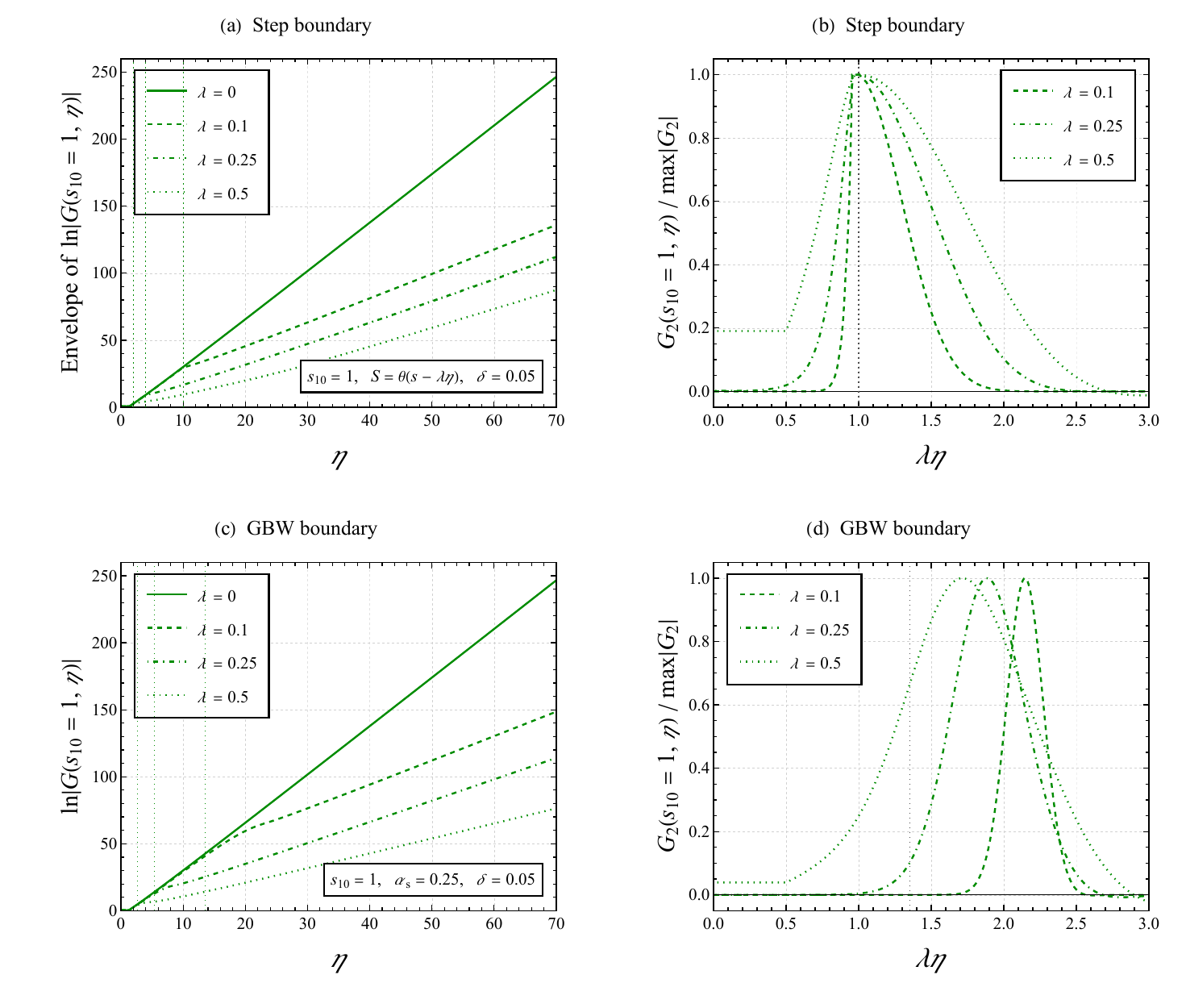}
\caption{Full numerical solution of \eq{eq:system} for a dipole of fixed size $s_{10}=1$, with $\alpha_s=0.25$ and $\delta=0.05$, for $\lambda=0$, $0.1$, $0.25$ and $0.5$. Top row: step boundary $S=\theta\Lb s-\lambda\eta\Rb$. Bottom row: GBW boundary, \eq{eq:GBW}. (a) Envelope of $\ln|G|$, i.e., maximum of $\ln|G|$ over $[\eta-4,\eta]$ (see text). (c) $\ln|G|$. In (a) and (c) the vertical dotted lines mark $\eta=s_{10}/\lambda$ (step) and $\eta=1.35\,s_{10}/\lambda$, where $S\Lb s_{10},\eta\Rb=1/2$ (GBW). (b), (d) $G_2$ with its sign, normalized to its maximum, as a function of $\lambda\eta$; the vertical dotted line marks $\lambda\eta=1$ (step) and $\lambda\eta=1.35$ (GBW).}
\label{fig:fixeddipole}
\end{figure*}

Figure~\ref{fig:fixeddipole} shows the full numerical solution of \eq{eq:system} for a dipole of fixed size $s_{10}=1$, with $\alpha_s=0.25$ and $\delta=0.05$, for the two forms of the saturation boundary introduced in Sec.~\ref{sec:boundary}: the step~(\ref{eq:step}) in panels (a) and (b), and the GBW form~(\ref{eq:GBW}) in panels (c) and (d).

\emph{Step boundary.} For $\eta<s_{10}/\lambda$ the dipole is dilute and $G$ grows with the KPS-CTT rate $3.66$. At $\eta=s_{10}/\lambda$ it enters the saturation region and the rate drops to $\omega\Lb \lambda\Rb$ [Fig.~\ref{fig:fixeddipole}(a)]. The amplitude $G_2$ grows until the dipole saturates and has a sharp maximum at $\lambda\eta\simeq s_{10}$, after which it decreases [Fig.~\ref{fig:fixeddipole}(b)]. The maximum is sharp because the step switches off the source of $G_2$, the IR term of \eq{eq:sysG2}, abruptly: before the maximum $G_2$ grows with the rate $3.66$ in $\eta$, that is, $3.66/\lambda$ in $\lambda\eta$, and at the maximum its slope drops to zero.

\emph{GBW boundary.} With the GBW form the transition has a width of order $\sqrt{\bas}$ in $s$, and $S\Lb s_{10},\eta\Rb=1/2$ at $\lambda\eta=s_{10}+\sqrt{\bas}\ln\Lb 4\ln2\Rb\simeq1.35$ (vertical dotted lines in panels (c) and (d)). The change of slope of $\ln|G|$ is then smooth and occurs somewhat later [Fig.~\ref{fig:fixeddipole}(c)]. The maximum of $G_2$ is rounded and moves to $\lambda\eta\simeq1.7$--$2.1$ [Fig.~\ref{fig:fixeddipole}(d)]: with a gradual boundary the source of $G_2$ fades out progressively, and $G_2$ starts to decrease only when the term $-\Lb 1-S\Rb\Gamma_2$, which requires saturated small daughters ($s_{21}>s_{10}$), becomes important. At $\eta=40$--$70$ the slopes of $\ln|G|$ with the GBW boundary are $1.82$, $1.59$ and $1.12$ for $\lambda=0.1$, $0.25$ and $0.5$, compared with $1.83$, $1.64$ and $1.41$ for the step and $1.90$, $1.73$ and $1.41$ for \eq{eq:omega_lambda}. The finite width of the GBW boundary slows down the approach to the asymptotic rate, in particular for large $\lambda$, as noted in Appendix~\ref{app:numerics}. The qualitative behavior, a reduced growth of $G$ and a $G_2$ that stops growing, is the same for both boundaries.

\emph{Envelope of $\ln|G|$.} With the step boundary and $\lambda=0.25$ and $0.5$, the amplitude $G$ changes sign at isolated values of $\eta$ inside the saturation region (for $\eta\simeq53$--$70$ and $\eta\simeq24$--$32$, respectively). At these points $\ln|G|$ has narrow downward spikes, of up to $11$ units. To display the growth rate we plot in panel (a) the envelope of $\ln|G|$, defined as the maximum of $\ln|G|$ over the interval $[\eta-4,\eta]$. The envelope removes the spikes and does not change the slopes. With the GBW boundary there are no sign changes in this range, and in panel (c) we plot $\ln|G|$ itself; its envelope would differ from it by less than $0.4$, and only near $\eta\simeq1$, where the Born term~(\ref{eq:Born_vars}) changes sign.

\begin{figure*}[tbp]
\centering
\includegraphics[width=0.9\textwidth]{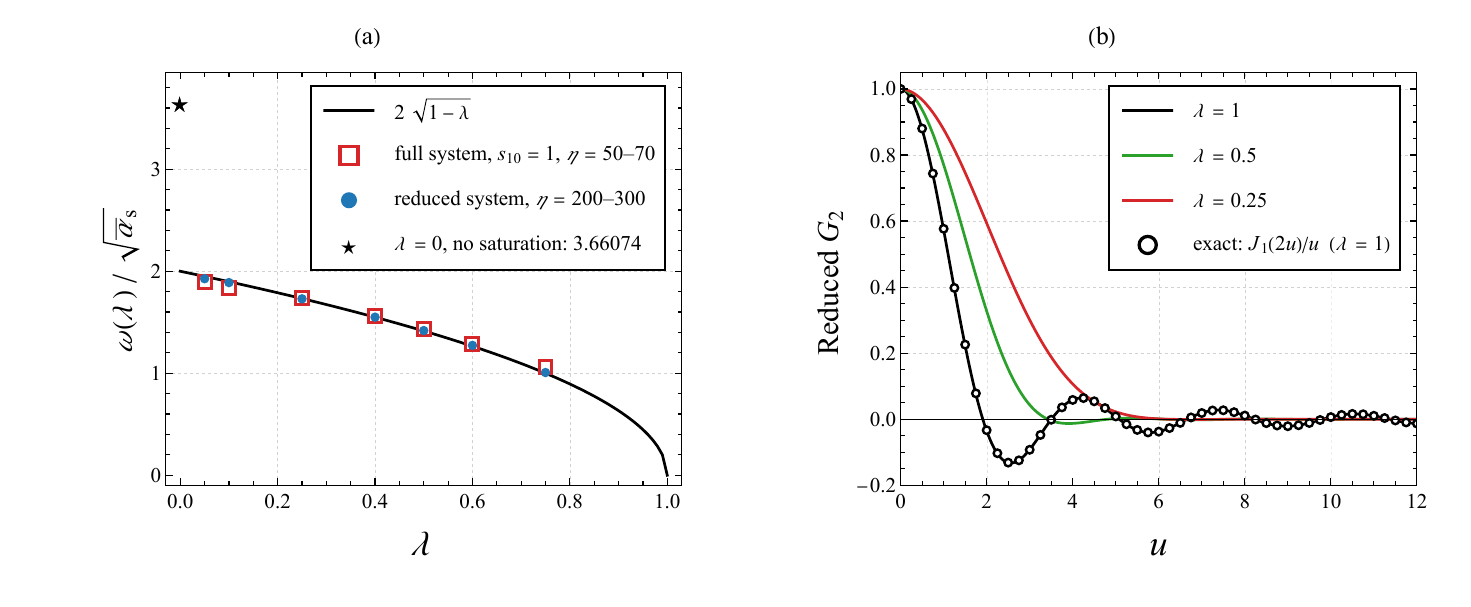}
\caption{(a) Growth rate of $G$ inside the saturation region, in units of $\sqrt{\bas}$: \eq{eq:omega_lambda} (line), reduced system (filled circles) and full system with $s_{10}=1$ (open squares), see Table~\ref{tab:omega}. The star at $\lambda=0$ is the rate without saturation ($S=1$), $\alpha_h=3.66074$; the curve does not reach it as $\lambda\to0^+$, because the limits $\lambda\to0$ and $\eta\to\infty$ do not commute (Sec.~\ref{sec:regime}). (b) $G_2$ from the reduced equation~(\ref{eq:G2red}) on the diagonal $U=V=u$, with its sign, for $\lambda=1$, $0.5$ and $0.25$. The open circles are the exact solution $J_1\Lb 2u\Rb/u$ for $\lambda=1$, \eq{eq:J0}.}
\label{fig:inside}
\end{figure*}

\emph{The limit $\lambda\to0$.} The rate inside the saturation region does not approach $3.66$ when $\lambda\to0$: for any $\lambda>0$ it is at most $2$. In Fig.~\ref{fig:inside}(a) this is the gap between the curve $2\sqrt{1-\lambda}$, which tends to $2$, and the star at $\lambda=0$, which marks the rate $3.66074$ without saturation. This is not a contradiction. The time at which a given dipole saturates, $\eta_{\mathrm{sat}}=s_{10}/\lambda$, goes to infinity as $\lambda\to0$, so the second regime is never reached. The limits $\lambda\to0$ and $\eta\to\infty$ do not commute, but for every $\lambda>0$ there is a well-defined change of regime at $\eta\simeq s_{10}/\lambda$.

For a dipole of fixed size the consequence is the following. With $\alpha_s=0.25$ ($\sqrt{\bas}=0.345$) and $\lambda=0.25$, the exponent of $G$ drops from $3.66\times0.345=1.26$ to $1.73\times0.345=0.60$ once the dipole is saturated. Observables that are dominated by dipoles of a fixed size, such as $\Delta G$, are therefore strongly affected by saturation. The quark helicity $\Delta\Sigma$, however, integrates over dipoles of all sizes down to $1/\sqrt{zs}$, and most of them are not saturated. Its behavior is determined in Sec.~\ref{sec:edge}.

\subsection{Asymptotic regime: the edge of the saturation region}
\label{sec:edge}

\begin{figure*}[tbp]
\centering
\includegraphics[width=0.9\textwidth]{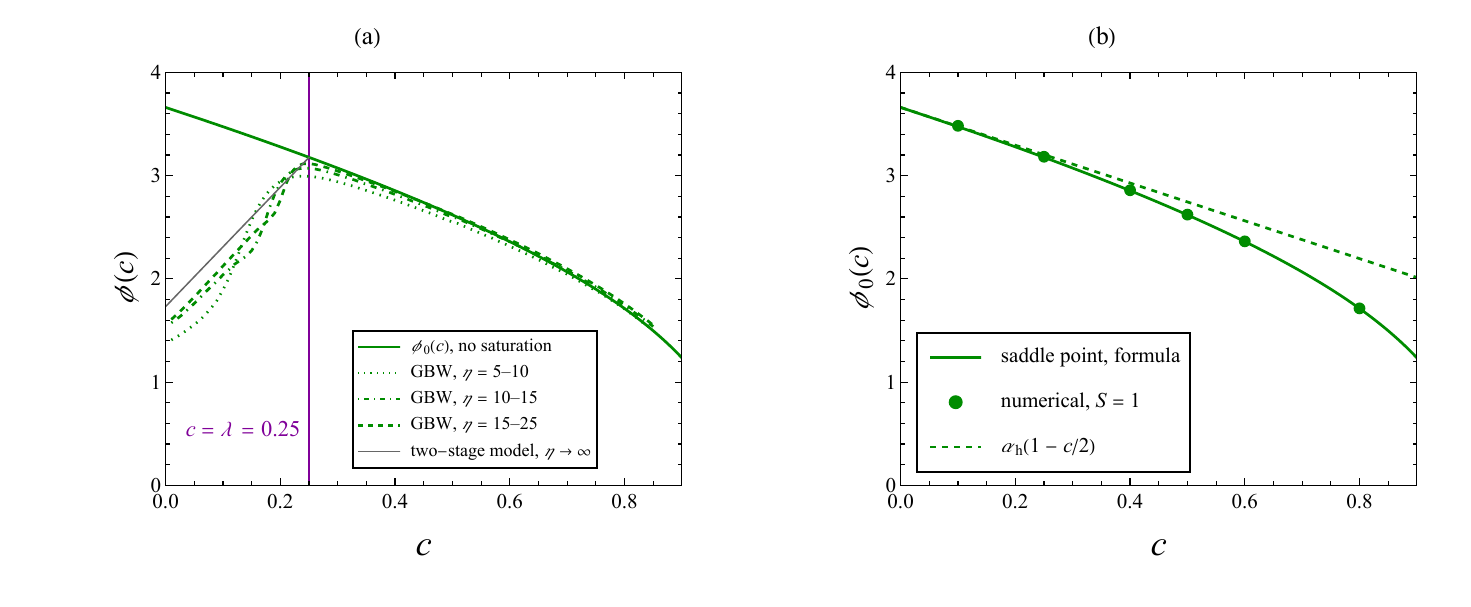}
\caption{(a) Growth rate $\phi\Lb c\Rb$ of $G$ along the rays $s=c\,\eta$, \eq{eq:rays}: without saturation, \eq{eq:phi0} (solid line), and with the GBW boundary, $\lambda=0.25$, measured at $\eta=5$--$10$ (dotted), $10$--$15$ (dot-dashed) and $15$--$25$ (dashed). The thin gray line is the two-stage model~(\ref{eq:twostage}), the limit $\eta\to\infty$, and the vertical line marks the edge of the saturation region, $c=\lambda$. At finite $\eta$ the curves are smooth, and their maximum at the edge sharpens as $\eta$ grows. (b) $\phi_0\Lb c\Rb$ from \eq{eq:phi0} (solid line), from the numerical solution with $S=1$ extrapolated to $\delta\to0$ (circles, Table~\ref{tab:phi0}), and the branch-point estimate $\alpha_h\Lb 1-c/2\Rb$ (dashed line).}
\label{fig:edge}
\end{figure*}

Consider the growth of $G\Lb s,\eta\Rb$ along the rays $s=c\,\eta$ in the $\Lb s,\eta\Rb$ plane (Fig.~\ref{fig:plane}),
\beq
\label{eq:rays}
G\Lb c\,\eta,\eta\Rb \sim e^{\phi\Lb c\Rb\,\eta}.
\eeq
The integration region of \eq{eq:DSigma_vars} contains, at large $\eta_s$, all rays $0\le c\le1$, so the exponent of $\Delta\Sigma$ is set by the maximum of $\phi\Lb c\Rb$. Without saturation, $\phi\Lb c\Rb$ is a decreasing function of $c$ with $\phi\Lb 0\Rb=\alpha_h$, and $\Delta\Sigma$ grows with the intercept $\alpha_h$. Figure~\ref{fig:edge}(a) shows $\phi\Lb c\Rb$ measured in our numerical solution with the GBW boundary~(\ref{eq:GBW}) with $\lambda=0.25$, which has an exactly straight saturation boundary, in three windows of rapidity, $\eta=5$--$10$, $10$--$15$ and $15$--$25$, together with the rate $\phi_0\Lb c\Rb$ without saturation [\eq{eq:phi0} below]. The rates are slopes of the envelope of $\ln|G\Lb c\,\eta,\eta\Rb|$, sampled every $0.01$ in $c$. For $c>\lambda$, that is, for dipoles that are not saturated, the rate is the same with and without saturation. For example, at $\eta=15$--$25$ we find $3.04$ against $3.06$ without saturation at $c=0.3$, and $2.63$ against $2.63$ at $c=0.5$. For $c<\lambda$ the rate drops, as found in Sec.~\ref{sec:asympt}. Therefore the maximum of $\phi\Lb c\Rb$ is at the edge of the saturation region, $c=\lambda$, and
\beq
\label{eq:aDSigma}
\alpha_{\Delta\Sigma} = \phi_0\Lb \lambda\Rb\,\sqrt{\bas},
\eeq
where $\phi_0\Lb c\Rb$ is the growth rate along the ray $c$ of the solution without saturation. For a target whose saturation scale grows with a slowly varying exponent, as in the rcBK evolution of Sec.~\ref{sec:rcbk}, $\lambda$ is the local slope of the boundary at the relevant rapidity.

\emph{Closed formula for $\phi_0$.} The function $\phi_0\Lb c\Rb$ can be obtained from the exact solution of Borden and Kovchegov~\cite{Borden:2023ugd}. In that solution the amplitudes are written as double inverse Mellin integrals over $\omega$ and $\gamma$ of $\exp[\omega\ln\Lb zsx_{10}^2\Rb+\gamma\ln\Lb 1/x_{10}^2\Lambda^2\Rb]$; for example, $G_2$ is the integral of this exponential times a function $G_{2\omega\gamma}$ whose denominator contains $\Lb \gamma-\gamma^-_\omega\Rb\Lb \gamma-\gamma^+_\omega\Rb$~\cite{Borden:2023ugd}. In the rescaled units $\omega\to\omega\sqrt{\bas}$, $\gamma\to\gamma\sqrt{\bas}$, the exponent is $\omega\Lb \eta-s_{10}\Rb+\gamma s_{10}$. Closing the $\gamma$ contour to the left picks the pole at $\gamma=\gamma^-_\omega$, where~\cite{Borden:2023ugd}
\beq
\label{eq:gammaminus}
\gamma^-_\omega = \frac{\omega}{2}\left[1-\sqrt{1-\frac{16}{\omega^2}\sqrt{1-\frac{4}{\omega^2}}}\right]
\eeq
is the resummed polarized anomalous dimension in rescaled units. The remaining integral has the form
\beq
\label{eq:omegaint}
G\Lb c\,\eta,\eta\Rb \sim \int\frac{d\omega}{2\pi i}\,e^{\eta\,f_c\Lb \omega\Rb}, \qquad f_c\Lb \omega\Rb = \Lb 1-c\Rb\,\omega + c\,\gamma^-_\omega,
\eeq
up to prefactors that do not affect the exponent. The rightmost singularity of the integrand is the branch point of $\gamma^-_\omega$ at $\omega=\alpha_h=3.66074$, where the inner square root of \eq{eq:gammaminus} vanishes; it determines the intercept for $c=0$~\cite{Borden:2023ugd}. For $c>0$ the asymptotics is instead given by a saddle point. Near the branch point $\gamma^-_\omega$ decreases as $-\sqrt{\omega-\alpha_h}$, with an infinite slope, so $f_c\Lb \omega\Rb$ always decreases to the right of $\alpha_h$; at large $\omega$ it increases because of the term $\Lb 1-c\Rb\omega$. Hence $f_c$ has a minimum on the real axis at some $\omega^\ast\Lb c\Rb>\alpha_h$, which is the saddle point of the integral~(\ref{eq:omegaint}) on a contour parallel to the imaginary axis. We obtain
\beq
\label{eq:phi0}
\boxed{\phi_0\Lb c\Rb = \min_{\omega\ge\alpha_h}\Big[\Lb 1-c\Rb\,\omega + c\,\gamma^-_\omega\Big].}
\eeq
For $c\to0$ the saddle point approaches the branch point and $\phi_0\to\alpha_h$; the expansion in $c$ is given in Appendix~\ref{app:saddle}. The naive estimate in which the integrand is evaluated at the branch point, $\phi_0\approx\alpha_h\Lb 1-c/2\Rb$, coincides with \eq{eq:phi0} only for $c\to0$ and overestimates the rate otherwise [Fig.~\ref{fig:edge}(b)]. Numerical values are listed in Table~\ref{tab:phi0}.

We have verified \eq{eq:phi0} against the numerical solution with $S=1$, measuring the rate along each ray at $\eta=20$--$40$ (with a logarithmic term in the fit) for two steps, and extrapolating linearly to $\delta\to0$ (Table~\ref{tab:phi0}). The extrapolated values agree with \eq{eq:phi0} to better than $0.25\%$ for all rays. The remaining difference is consistent with the subleading terms of the fit at finite $\eta$.

\begin{table}[tbp]
\caption{Growth rate $\phi_0\Lb c\Rb$ of $G$ along the ray $s=c\,\eta$ without saturation: numerical solution for two steps $\delta$, linear extrapolation $\delta\to0$, and \eq{eq:phi0} with the position $\omega^\ast$ of the saddle point.}
\label{tab:phi0}
\begin{ruledtabular}
\begin{tabular}{ccccccc}
$c$ & $\delta=0.05$ & $\delta=0.025$ & $\delta\to0$ & \eq{eq:phi0} & diff. & $\omega^\ast$ \\ \hline
0.1 & 3.4840 & 3.4833 & 3.4826 & 3.4739 & 0.25\% & 3.665 \\
0.25 & 3.1973 & 3.1901 & 3.1829 & 3.1772 & 0.18\% & 3.691 \\
0.4 & 2.8867 & 2.8726 & 2.8585 & 2.8545 & 0.14\% & 3.759 \\
0.5 & 2.6602 & 2.6414 & 2.6226 & 2.6193 & 0.13\% & 3.847 \\
0.6 & 2.4111 & 2.3873 & 2.3635 & 2.3608 & 0.11\% & 4.002 \\
0.8 & 1.7857 & 1.7500 & 1.7143 & 1.7134 & 0.05\% & 4.902
\end{tabular}
\end{ruledtabular}
\end{table}

\emph{The kink at $c=\lambda$ and its smoothing.} Asymptotically, $\phi\Lb c\Rb$ has a kink at $c=\lambda$. Its shape for $c<\lambda$ can be understood with a simple two-stage approximation: a dipole on the ray $c<\lambda$ is dilute until $\eta_e=c\,\eta/\lambda$, when it reaches the boundary, and grows with the rate $\phi_0\Lb \lambda\Rb$ up to that point; afterwards it grows with the rate $\omega\Lb \lambda\Rb=2\sqrt{1-\lambda}$ of \eq{eq:omega_lambda}. This gives
\beq
\label{eq:twostage}
\phi\Lb c\Rb \simeq \frac{c}{\lambda}\,\phi_0\Lb \lambda\Rb + \Big(1-\frac{c}{\lambda}\Big)\,2\sqrt{1-\lambda}, \qquad c<\lambda,
\eeq
which is linear in $c$, interpolates between $2\sqrt{1-\lambda}$ at $c=0$ and $\phi_0\Lb \lambda\Rb$ at $c=\lambda$, and is shown as a thin gray line in Fig.~\ref{fig:edge}(a).

\emph{The kink is a property of the limit $\eta\to\infty$.} The two-stage model assumes that the dipole changes regime instantaneously, and it describes the limit $\eta\to\infty$, in which $\phi\Lb c\Rb=\lim_{\eta\to\infty}\eta^{-1}\ln G\Lb c\,\eta,\eta\Rb$ has a kink at $c=\lambda$: the function is continuous there, but its slope jumps. At any finite $\eta$ the transition is smooth, because the boundary has a width of order $\sqrt{\bas}$ in $s$ (Sec.~\ref{sec:boundary}) and because a dipole close to the edge spends a finite range of rapidity near the boundary. This is seen in Fig.~\ref{fig:edge}(a): at $\eta=5$--$10$ the maximum is broad and rounded, $\phi=2.99$ at $c=0.24$ with $\phi=2.94$ at both $c=0.2$ and $c=0.3$; at $\eta=15$--$25$ it is narrower, $\phi=3.12$ at $c=0.25$ with $2.76$ at $c=0.2$ and $3.04$ at $c=0.3$, and it approaches the shape of the two-stage model. In all three windows the maximum lies at the edge of the saturation region, $c\simeq0.24$--$0.25$, and its value approaches $\phi_0\Lb \lambda\Rb=3.18$ as $\eta$ grows. This is the property that enters \eq{eq:aDSigma}, which only involves the value at the maximum; the kink itself plays no role. The situation is analogous to the rate functions of statistical mechanics, which can have kinks in the thermodynamic limit while every finite system is smooth.

\begin{table}[tbp]
\caption{Values of \eq{eq:phi0}, of the exponent~(\ref{eq:aDSigma}) for $\alpha_s=0.25$, and of the convergence threshold~(\ref{eq:astar}).}
\label{tab:astar}
\begin{ruledtabular}
\begin{tabular}{lccccc}
$\lambda$ & 0 & 0.2 & 0.25 & 0.3 & 0.4 \\ \hline
$\phi_0\Lb \lambda\Rb$ & 3.6607 & 3.2786 & 3.1772 & 3.0730 & 2.8545 \\
$\alpha_{\Delta\Sigma}$ ($\alpha_s=0.25$) & 1.265 & 1.133 & 1.098 & 1.062 & 0.986 \\
$\alpha_s^\ast\Lb \lambda\Rb$ & 0.156 & 0.195 & 0.208 & 0.222 & 0.257
\end{tabular}
\end{ruledtabular}
\end{table}

\subsection{Asymptotic convergence of the helicity integrals}
\label{sec:convergence}

The integral $\int_0dx\,\Delta\Sigma\Lb x,Q^2\Rb$ converges at small $x$ if $\alpha_{\Delta\Sigma}<1$. With \eq{eq:aDSigma} this gives
\beq
\label{eq:astar}
\alpha_s < \alpha_s^\ast\Lb \lambda\Rb = \frac{2\pi}{N_c\,\phi_0^2\Lb \lambda\Rb}, 
\eeq
listed in Table~\ref{tab:astar}. Without saturation, $\alpha_s^\ast\Lb 0\Rb=0.156$ [\eq{eq:threshold0}]. With $\lambda=0.25$ the threshold rises to $0.208$, and with $\lambda=0.3$ to $0.222$. For $\lambda\approx0.3$, the value found in the HERA data~\cite{Golec-Biernat:1998zce,Stasto:2000er} and in our rcBK solution in the experimental range (Sec.~\ref{sec:rcbk}), the threshold is $\alpha_s^\ast\approx0.22$. Equation~(\ref{eq:astar}) is a property of the asymptotic regime of the equations for a boundary of constant slope; it is not a statement about the range of $x$ accessible to experiment, where the effective exponents are those of Sec.~\ref{sec:observables}.

\emph{The gluon helicity.} The same analysis applies to $\Delta G$, \eq{eq:DG_DLA}, with a different outcome. $\Delta G$ is given by $G_2$ at the fixed size $1/Q$, and $G_2$ is fed by the larger dipoles through the IR term of \eq{eq:sysG2}. Once these dipoles and the dipole $1/Q$ itself are inside the saturation region, $G_2$ stops growing and oscillates with a decreasing amplitude (Sec.~\ref{sec:G2inside} and Fig.~\ref{fig:fixeddipole}). Hence, in the asymptotic regime, $\Delta G$ reaches a maximum and decreases, and the integral $\int_0dx\,\Delta G$ computed with the asymptotic solution converges for any coupling. This is a property of the asymptotic regime. For the target of Sec.~\ref{sec:numerics} it is not reached in the experimental range, where $\Delta G$ still grows, with a reduced exponent (Sec.~\ref{sec:observables}).

\section{Numerical solution with running-coupling BK evolution}
\label{sec:numerics}

We now solve the system~(\ref{eq:system}) numerically with a realistic $S$-matrix and compute the helicity distributions~(\ref{eq:DSigma}) and (\ref{eq:DG_DLA}) in the range of $x$ that is accessible to experiment, $10^{-2}\ge x\ge10^{-6}$. The Electron-Ion Collider will reach $x\sim10^{-4}$ in polarized DIS~\cite{Accardi:2012qut,AbdulKhalek:2021gbh}, and HERA reached $x\sim10^{-6}$ in unpolarized DIS~\cite{H1:2015ubc}; we refer to these values as the EIC reach and the HERA reach. The numerical method is described in Appendix~\ref{app:numerics}. Throughout this section the helicity evolution is at fixed coupling, as in Refs.~\cite{Cougoulic:2022gbk,Borden:2023ugd}, with $N_c=3$ and, unless stated otherwise, $\alpha_s=0.25$, i.e., $\sqrt{\bas}=0.3455$. The numerical data of all the figures are provided in the Supplemental Material~\cite{SupplementalMaterial}.

\subsection{Validation: the limit $S=1$}
\label{sec:validation}

\emph{Purpose.} The numerical results of this paper rest on the numerical solution of the system~(\ref{eq:system}). Before saturation is switched on, the numerical solution must reproduce the known solution of the helicity evolution without saturation. This is a test of the limit: for $S=1$ the system~(\ref{eq:system}) reduces exactly to the equations of Ref.~\cite{Cougoulic:2022gbk}, whose asymptotic growth is known analytically, \eq{eq:alphah}~\cite{Borden:2023ugd}. If it did not reproduce $\alpha_h=3.66074\sqrt{\bas}$ for $S=1$, none of the results with saturation could be trusted. The test checks at the same time the discretization, the treatment of the neighbor amplitudes and the inhomogeneous terms, and the method used to extract growth rates, which is the same method that we use with saturation in Secs.~\ref{sec:asympt} and \ref{sec:edge}.

\emph{Two independent solutions.} We have solved the problem in two independent ways. The first solves only the KPS-CTT equations with $S=1$, using the discretization of Ref.~\cite{Cougoulic:2022gbk} and the closed form of $\Gamma_2$ of Appendix~\ref{app:numerics}. The second solves the system~(\ref{eq:system}) for an arbitrary function $S\Lb s,\eta\Rb$, including the new terms proportional to $1-S$. For $S=1$ the two solutions agree on the whole grid within round-off. This shows that the new terms vanish when they should, and that no error was introduced when the $S$-matrices were added. All results with saturation are obtained with the second solution.

\begin{figure*}[tbp]
\centering
\includegraphics[width=0.9\textwidth]{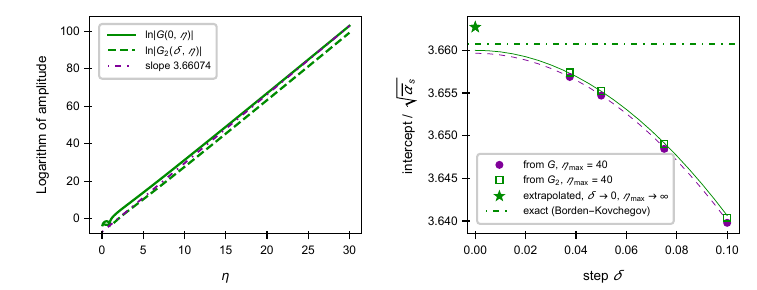}
\caption{Validation of the numerical solution without saturation ($S=1$). (a) $\ln|G\Lb 0,\eta\Rb|$ (solid line) and $\ln|G_2\Lb \delta,\eta\Rb|$ (dashed line) for the step $\delta=0.05$; the dot-dashed line has the exact slope $3.66074$~\cite{Borden:2023ugd}. Both amplitudes grow with this slope; the same quantities obtained in Ref.~\cite{Cougoulic:2022gbk} are shown in Fig.~7 of that reference. The kink at $\eta\simeq1$ is a change of sign of the Born term~(\ref{eq:Born_vars}). (b) Intercept, in units of $\sqrt{\bas}$, extracted from $G$ (filled circles) and from $G_2$ (open squares) with $\eta_{\max}=40$, as a function of the step $\delta$; each point is an independent solution. The thin dashed ($G$) and solid ($G_2$) lines are the fit~(\ref{eq:extrapfit}) at $\eta_{\max}=40$, the star is its extrapolation to $\delta\to0$, $\eta_{\max}\to\infty$, $3.6628$, and the dot-dashed line is the exact value $3.66074$. The extrapolated intercept differs from the exact one by $0.06\%$.}
\label{fig:validation}
\end{figure*}

\emph{Extraction of the intercept.} At large $\eta$ the amplitudes grow as $G\Lb 0,\eta\Rb\sim e^{\alpha_h\eta}$, up to a power of $\eta$ in the prefactor, so that $\ln|G\Lb 0,\eta\Rb|$ is asymptotically a straight line whose slope is the intercept in units of $\sqrt{\bas}$ [Fig.~\ref{fig:validation}(a); see also Fig.~7 of Ref.~\cite{Cougoulic:2022gbk}]. We extract it by fitting $\ln|G|=a\,\eta+b\ln\eta+c$ in the upper half of the range $0\le\eta\le\eta_{\max}$; the term $b\ln\eta$ absorbs the power-law prefactor. For $G_2$ we use the first point of the grid, $s_{10}=\delta$. In the continuum equations $G_2\Lb 0,\eta\Rb=0$, since the IR integral in \eq{eq:sysG2} is empty at $s_{10}=0$ and the Born term~(\ref{eq:Born_vars}) is proportional to $s_{10}$; with our grid the amplitude at the first point plays the role of $G_2$ at small dipole size, and its slope gives the same intercept. (In Fig.~7 of Ref.~\cite{Cougoulic:2022gbk} the corresponding curve is labeled $G_2\Lb 0,\eta\Rb$, with the grid conventions of that reference.) The kink of $\ln|G|$ near $\eta\simeq1$ in Fig.~\ref{fig:validation}(a) is a change of sign of the Born term, $G^{\Lb 0\Rb}\Lb 0,\eta\Rb\propto\Lb C_F-2\Rb\eta<0$, which is reversed by the evolution; it does not affect the slope.

\emph{Why an extrapolation is needed.} A single run does not give the exact intercept, for two reasons. First, the discretization with step $\delta$ introduces an error that vanishes as $\delta\to0$; with a coarse grid the growth is slightly underestimated. Second, at finite $\eta_{\max}$ the fit is affected by subleading terms that vanish as $\eta_{\max}\to\infty$. Both effects are visible in Fig.~\ref{fig:validation}(b): for $\eta_{\max}=40$ the intercept measured with $G$ (circles) and with $G_2$ (squares) increases monotonically from about $3.640$ at $\delta=0.1$ to about $3.657$ at $\delta=0.0375$. We therefore computed the intercept for $\delta=0.1$, $0.075$, $0.05$, $0.0375$ and $\eta_{\max}=20$, $30$, $40$, twelve runs in total, and fitted all of them with
\beq
\label{eq:extrapfit}
a\Lb \delta,\eta_{\max}\Rb = c_0 + c_1\delta + c_2\delta^2 + \frac{c_3}{\eta_{\max}} .
\eeq
The constant $c_0$ is the intercept extrapolated to $\delta\to0$ and $\eta_{\max}\to\infty$. The same procedure was used in Ref.~\cite{Cougoulic:2022gbk}. The thin lines in Fig.~\ref{fig:validation}(b) show the fit~(\ref{eq:extrapfit}) at $\eta_{\max}=40$, and the star its extrapolation. We obtain
\beq
\label{eq:validation}
\begin{split}
\alpha_h &= 3.6628\,\sqrt{\bas} \quad \Lb \text{from } G\Rb, \\
\alpha_{h,2} &= 3.6615\,\sqrt{\bas} \quad \Lb \text{from } G_2\Rb,
\end{split}
\eeq
to be compared with the numerical values $\Lb 3.661\pm0.006\Rb\sqrt{\bas}$ and $\Lb 3.660\pm0.009\Rb\sqrt{\bas}$ of Ref.~\cite{Cougoulic:2022gbk} and with the exact result~(\ref{eq:alphah}), $3.66074\sqrt{\bas}$ (dot-dashed line). The differences from the exact value are $0.06\%$ and $0.02\%$, respectively. The two amplitudes give the same intercept, as they must, since $G$ and $G_2$ are coupled and grow with a common exponent~\cite{Cougoulic:2022gbk,Borden:2023ugd}.

\emph{What this implies for the results with saturation.} The test is a test of the numerical method; it is performed at large $\eta$, beyond the experimental range, because the intercept is defined there. It fixes the numerical accuracy of all growth rates quoted below. With the steps used in this paper, $\delta=0.05$--$0.1$, an intercept is underestimated by at most $0.6\%$ at $\eta_{\max}=40$, which is much smaller than the effects of saturation that we discuss: inside the saturation region the rate of $G$ falls from $3.66\sqrt{\bas}$ to $2\sqrt{1-\lambda}\,\sqrt{\bas}\le2\sqrt{\bas}$ [\eq{eq:omega_lambda}], and the exponents of the observables change at the level of tens of percent. Where a more accurate value is needed, as for the growth rates along rays in Table~\ref{tab:phi0}, we use the same extrapolation in $\delta$.

\subsection{The unpolarized $S$-matrix from running-coupling BK}
\label{sec:rcbk}

For the unpolarized $S$-matrix we use the solution of the BK equation with running coupling (rcBK) for a target that does not depend on the impact parameter. Writing $S=1-N$, the equation reads~\cite{Balitsky:2006wa,Albacete:2007yr}
\beq
\label{eq:rcBK}
\begin{split}
\frac{\partial N\Lb r,Y\Rb}{\partial Y} = \int d^2r_1\,K\Lb r,r_1,r_2\Rb\Big[&N\Lb r_1\Rb+N\Lb r_2\Rb-N\Lb r\Rb \\
&-N\Lb r_1\Rb N\Lb r_2\Rb\Big],
\end{split}
\eeq
with ${\un r}_2={\un r}-{\un r}_1$ and Balitsky's prescription for the running coupling~\cite{Balitsky:2006wa},
\beq
\label{eq:Kbal}
\begin{split}
K = \frac{N_c\,\alpha_s\Lb r^2\Rb}{2\pi^2}\Bigg[&\frac{r^2}{r_1^2r_2^2} + \frac{1}{r_1^2}\left(\frac{\alpha_s\Lb r_1^2\Rb}{\alpha_s\Lb r_2^2\Rb}-1\right) \\
&+ \frac{1}{r_2^2}\left(\frac{\alpha_s\Lb r_2^2\Rb}{\alpha_s\Lb r_1^2\Rb}-1\right)\Bigg].
\end{split}
\eeq
The one-loop coupling in coordinate space is $\alpha_s\Lb r^2\Rb=4\pi/[\beta_0\ln\Lb 4e^{-2\gamma_E}/r^2\Lambda_{\mathrm{QCD}}^2\Rb]$, with $\beta_0=\Lb 11N_c-2N_f\Rb/3$, $N_f=3$ and $\Lambda_{\mathrm{QCD}}=0.241$~GeV, frozen at $\alpha_s=0.7$ in the infrared~\cite{Albacete:2007yr,Albacete:2009fh}. The initial condition is of the McLerran--Venugopalan type~\cite{McLerran:1993ni,McLerran:1993ka},
\beq
\label{eq:MV}
N\Lb r,Y=0\Rb = 1-\exp\left[-\frac{r^2Q_{s0}^2}{4}\ln\left(\frac{1}{r\Lambda_{\mathrm{QCD}}}+e\right)\right],
\eeq
where $Q_{s0}$ is the initial saturation scale, given below for each scheme. We define the saturation scale by $N\Lb r=\sqrt2/Q_s,Y\Rb=1/2$.

As a check, we have solved the equation with a fixed coupling $\alpha_s=0.25$. The velocity $d\ln Q_s^2/dY$ grows as $0.77$, $0.93$, $1.00$ and $1.04$ at $Y=6$, $10$, $14$ and $18$, and slowly approaches the asymptotic value $4.88\,\alpha_sN_c/\pi=1.165$ of the leading-order BK equation~\cite{Mueller:2002zm}, as expected.

\emph{Starting points of the evolution: three schemes.} Both evolutions, that of the polarized amplitudes and that of the target, start at some value $x_0$ below which the small-$x$ resummation is applied; above it the amplitudes are given by their initial conditions. The value of $x_0$ is not fixed by the theory. The analyses of polarized data with small-$x$ helicity evolution start it at $x_0=0.1$~\cite{Adamiak:2021ppq,Adamiak:2023yhz}, while the descriptions of the HERA data with rcBK start the evolution of the target at $x_0=0.01$~\cite{Albacete:2009fh,Albacete:2010sy,Lappi:2013zma}. In terms of the variables~(\ref{eq:vars}), starting the helicity evolution at $x_0$ amounts to replacing $1/x$ by $x_0/x$ in the relation between $zs$ and $x$. To expose the dependence of the results on this choice, we consider three schemes:
\begin{itemize}
\item[A:] helicity and target evolve from $x_0=0.1$;
\item[B:] helicity and target evolve from $x_0=0.01$;
\item[C:] the helicity evolves from $x_0=0.1$, as in Refs.~\cite{Adamiak:2021ppq,Adamiak:2023yhz}, and the target from $x=0.01$, as in Refs.~\cite{Albacete:2009fh,Albacete:2010sy}; for $0.1>x>0.01$ the target is given by the initial condition~(\ref{eq:MV}).
\end{itemize}
Scheme C combines the standard choices of the two communities, and we use it as our central scheme. In schemes B and C the target is the same, with $Q_{s0}^2=0.2$~GeV$^2$ at $x=0.01$. In scheme A we take $Q_{s0}^2=0.036$~GeV$^2$ at $x=0.1$. This value is chosen so that the saturation scale of scheme A coincides with that of schemes B and C in the whole range $10^{-2}\ge x\ge10^{-6}$, to better than $4\%$. The three schemes therefore describe the same proton in the experimental range, and they differ only in where the evolution starts.

The saturation scale of our target is compared with that of the GBW fit to the HERA data, $Q_{s,\mathrm{GBW}}^2=\Lb 3\times10^{-4}/x\Rb^{0.288}$~GeV$^2$~\cite{Golec-Biernat:1998zce}, in Table~\ref{tab:Qs}. The GBW model defines the saturation scale through $N=1-\exp\Lb -r^2Q_s^2/4\Rb$; with our definition $N\Lb \sqrt2/Q_s\Rb=1/2$ its saturation scale is multiplied by $2/\Lb 4\ln2\Rb=0.72$, which is the value listed in the table. Our $Q_s^2$ is about $1.4$--$1.5$ times the GBW value in the range $10^{-4}\ge x\ge10^{-6}$, a difference within the spread of the saturation models fitted to HERA~\cite{Golec-Biernat:1998zce,Iancu:2003ge,Kowalski:2003hm,Rezaeian:2012ji}; the dependence of our results on the infrared region, where this difference matters most, is studied in Sec.~\ref{sec:schemes}.

\begin{table}[tbp]
\caption{Saturation scale $Q_s^2$ (GeV$^2$), defined by $N\Lb r=\sqrt2/Q_s\Rb=1/2$, in the three schemes and in the GBW fit to HERA~\cite{Golec-Biernat:1998zce} converted to the same definition.}
\label{tab:Qs}
\begin{ruledtabular}
\begin{tabular}{lcccccc}
$x$ & $10^{-1}$ & $10^{-2}$ & $10^{-3}$ & $10^{-4}$ & $10^{-5}$ & $10^{-6}$ \\ \hline
Scheme A & 0.031 & 0.20 & 0.69 & 1.42 & 2.77 & 5.45 \\
Schemes B, C & --- & 0.20 & 0.66 & 1.42 & 2.83 & 5.62 \\
GBW (HERA) & --- & 0.26 & 0.51 & 0.99 & 1.92 & 3.72
\end{tabular}
\end{ruledtabular}
\end{table}

\begin{figure*}[tbp]
\centering
\includegraphics[width=\textwidth]{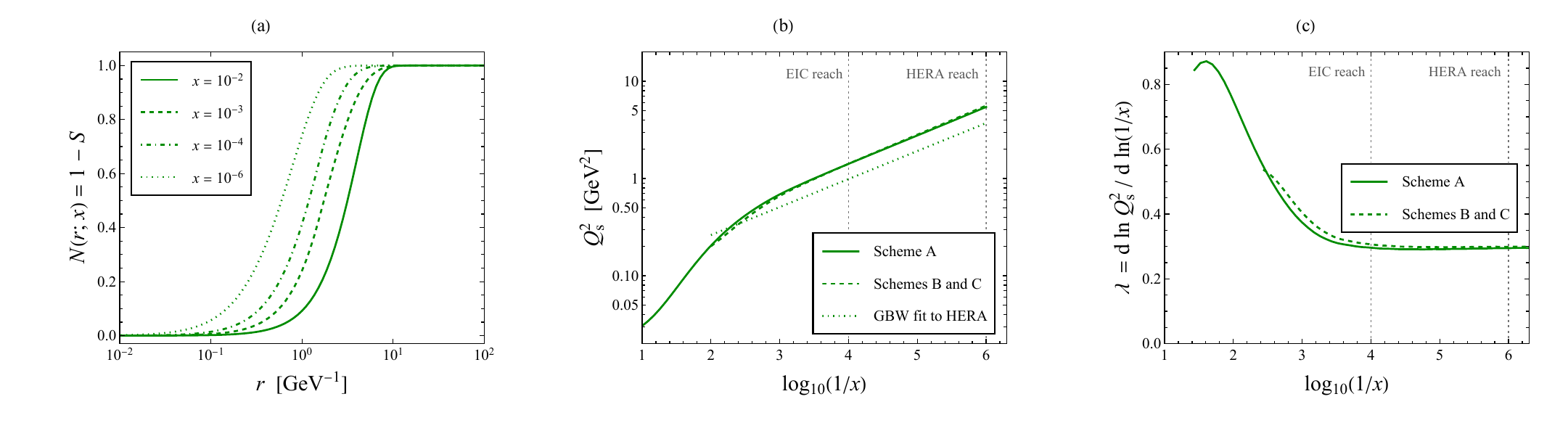}
\caption{The target in the experimental range. (a) Dipole amplitude $N\Lb r;x\Rb=1-S$ of schemes B and C, from the rcBK equation~(\ref{eq:rcBK}) with the initial condition~(\ref{eq:MV}) at $x=0.01$, for $x=10^{-2}$, $10^{-3}$, $10^{-4}$ and $10^{-6}$. (b) Saturation scale in scheme A (solid), in schemes B and C (dashed), and in the GBW fit to the HERA data converted to our definition (dotted), Table~\ref{tab:Qs}. (c) Local exponent~(\ref{eq:lamloc}) in scheme A (solid) and in schemes B and C (dashed), averaged over $\Delta Y\approx1.8$ to remove small oscillations produced by the discrete grid in $r$; it is shown where the target evolves, from $x=0.1$ in scheme A and from $x=0.01$ in schemes B and C. In (b) and (c) the vertical lines mark the EIC reach, $x\sim10^{-4}$, and the HERA reach, $x\sim10^{-6}$.}
\label{fig:rcbk}
\end{figure*}

\emph{The evolution of the target.} Figure~\ref{fig:rcbk}(a) shows the dipole amplitude $N=1-S$ of schemes B and C at $x=10^{-2}$ (the initial condition), $10^{-3}$, $10^{-4}$ and $10^{-6}$: as $x$ decreases, the amplitude moves to smaller dipoles. Figure~\ref{fig:rcbk}(b) shows $Q_s^2\Lb x\Rb$ in the three schemes and in the GBW fit, and Fig.~\ref{fig:rcbk}(c) the local exponent
\beq
\label{eq:lamloc}
\lambda\Lb x\Rb=\frac{d\ln Q_s^2}{d\ln\Lb 1/x\Rb} .
\eeq
In the experimental range the exponent is $\lambda\simeq0.30$ in all schemes ($0.31$ at $x=10^{-4}$ and $0.30$ at $x=10^{-6}$), in agreement with the value $\lambda\approx0.29$ of the saturation models fitted to the HERA data~\cite{Golec-Biernat:1998zce,Golec-Biernat:1999qor,Bartels:2002cj,Iancu:2003ge} and of geometric scaling~\cite{Stasto:2000er}. At the beginning of the evolution $\lambda$ first increases and then decreases. The increase occurs because the initial condition~(\ref{eq:MV}) is not a solution of the rcBK equation, which approaches a traveling-wave front~\cite{Munier:2003vc,Mueller:2002zm}: in the first units of rapidity the evolution builds up the evolved shape of the amplitude at small $r$, and the point $N=1/2$ moves quickly. The subsequent decrease is an effect of the running coupling: $\lambda$ is roughly proportional to $\alpha_s$ at the scale $Q_s$, which decreases as $Q_s$ grows. In scheme A the initial increase is larger, up to $\lambda\approx0.9$, because the evolution starts from a dilute proton, $Q_{s0}^2=0.036$~GeV$^2$, for which the relevant dipoles are in the region where the coupling is frozen at $0.7$. Since $Q_s$ is obtained by interpolating the solution between the points of the grid in $r$, the numerical derivative~(\ref{eq:lamloc}) shows small oscillations, of about $\pm0.008$; in Fig.~\ref{fig:rcbk}(c) they are removed by a moving average over $\Delta Y\approx1.8$.

\emph{The $S$-matrix in the helicity evolution.} To use the rcBK solution in the system~(\ref{eq:system}) we identify the rapidity of the helicity evolution, measured from $x_0$, with $\eta/\sqrt{\bas}=\ln\Lb zs/\Lambda^2\Rb$, and the dipole size with $r=e^{-s/2\sqrt{\bas}}/\Lambda$:
\beq
\label{eq:Smap}
\begin{split}
S\Lb s,\eta\Rb &= 1-N\big(r,Y\big), \qquad r=\frac{1}{\Lambda}\,e^{-s/2\sqrt{\bas}}, \\
Y &= \max\Big[0,\ \frac{\eta}{\sqrt{\bas}}-\ln\frac{x_0}{x_T}\Big],
\end{split}
\eeq
where $Y$ is the rapidity of the target evolution and $x_T$ the value of $x$ at which it starts: $x_T=x_0$ in schemes A and B, and $x_T=0.01$, $x_0=0.1$ in scheme C. Our default choice of the infrared scale is $\Lambda^2=0.2$~GeV$^2$, so that $s=0$ corresponds to dipoles of the size of the initial saturation scale of schemes B and C; to estimate the sensitivity to this choice we also use $\Lambda^2=0.05$ and $0.8$~GeV$^2$. The helicity distributions are computed from \eq{eq:DG_DLA} and from \eq{eq:DSigma}, which in the variables~(\ref{eq:vars}) reads
\beq
\label{eq:DSigma_vars}
\Delta\Sigma\Lb x,Q^2\Rb \propto -\int\limits_0^{\eta_s}d\eta'\!\!\int\limits_{\max[0,\,\eta'-y]}^{\eta'}\!\!ds_{10}\,\big[G+2G_2\big]\Lb s_{10},\eta'\Rb,
\eeq
with $\eta_s=\sqrt{\bas}\ln\Lb x_0Q^2/x\Lambda^2\Rb$ and $y=\sqrt{\bas}\ln\Lb x_0/x\Rb$, and we use $Q\approx G$. Since the normalization of the Born terms~(\ref{eq:Born_vars}) is a common factor, we only discuss ratios and exponents.

\begin{figure}[tbp]
\centering
\includegraphics[width=\columnwidth]{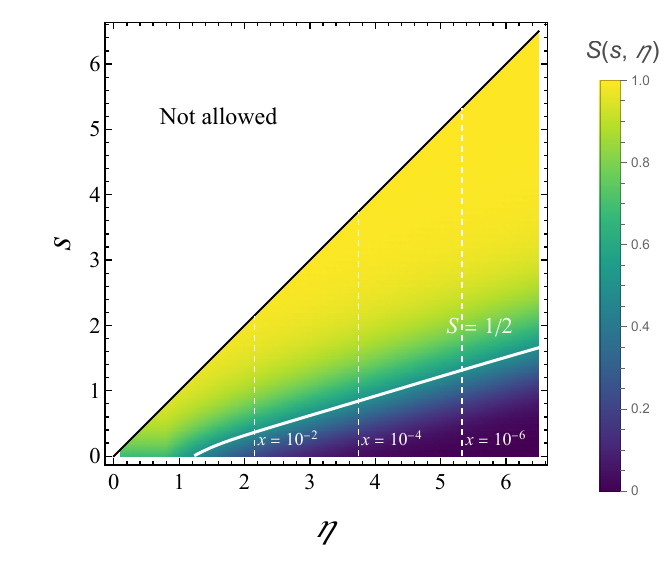}
\caption{Unpolarized $S$-matrix in the $\Lb s,\eta\Rb$ plane of the helicity evolution, \eq{eq:Smap}, in scheme C, for $\alpha_s=0.25$ and $\Lambda^2=0.2$~GeV$^2$. The white line is $S=1/2$. The vertical dashed lines mark the values of $\eta_s$ that correspond to $x=10^{-2}$, $10^{-4}$ and $10^{-6}$ at $Q^2=10$~GeV$^2$. The region $s>\eta$ is not allowed.}
\label{fig:Smap}
\end{figure}

The resulting $S\Lb s,\eta\Rb$ in scheme C is shown in Fig.~\ref{fig:Smap}. At $\eta=0$, that is, at $x_0=0.1$, the dipoles of size $1/\Lambda$ ($s=0$) are already partly saturated by the initial condition, $S=0.68$. The target does not evolve until $\eta=\sqrt{\bas}\ln10=0.80$, which corresponds to $x=0.01$; afterwards the saturation region grows toward smaller dipoles, and the line $S=1/2$ reaches $s=0.32$, $0.92$ and $1.51$ at $\eta=2$, $4$ and $6$. For $Q^2=10$~GeV$^2$ the values $x=10^{-2}$, $10^{-4}$ and $10^{-6}$ correspond to $\eta_s=2.15$, $3.74$ and $5.33$. In the experimental range the saturation boundary is therefore close to the origin of the $\Lb s,\eta\Rb$ plane, and the asymptotic regime of Sec.~\ref{sec:asympt}, which requires $\eta\gg s/\lambda$, is not reached for the dipoles of size $1/Q$.

\subsection{The helicity distributions in the experimental range}
\label{sec:observables}

\begin{figure*}[tbp]
\centering
\includegraphics[width=0.9\textwidth]{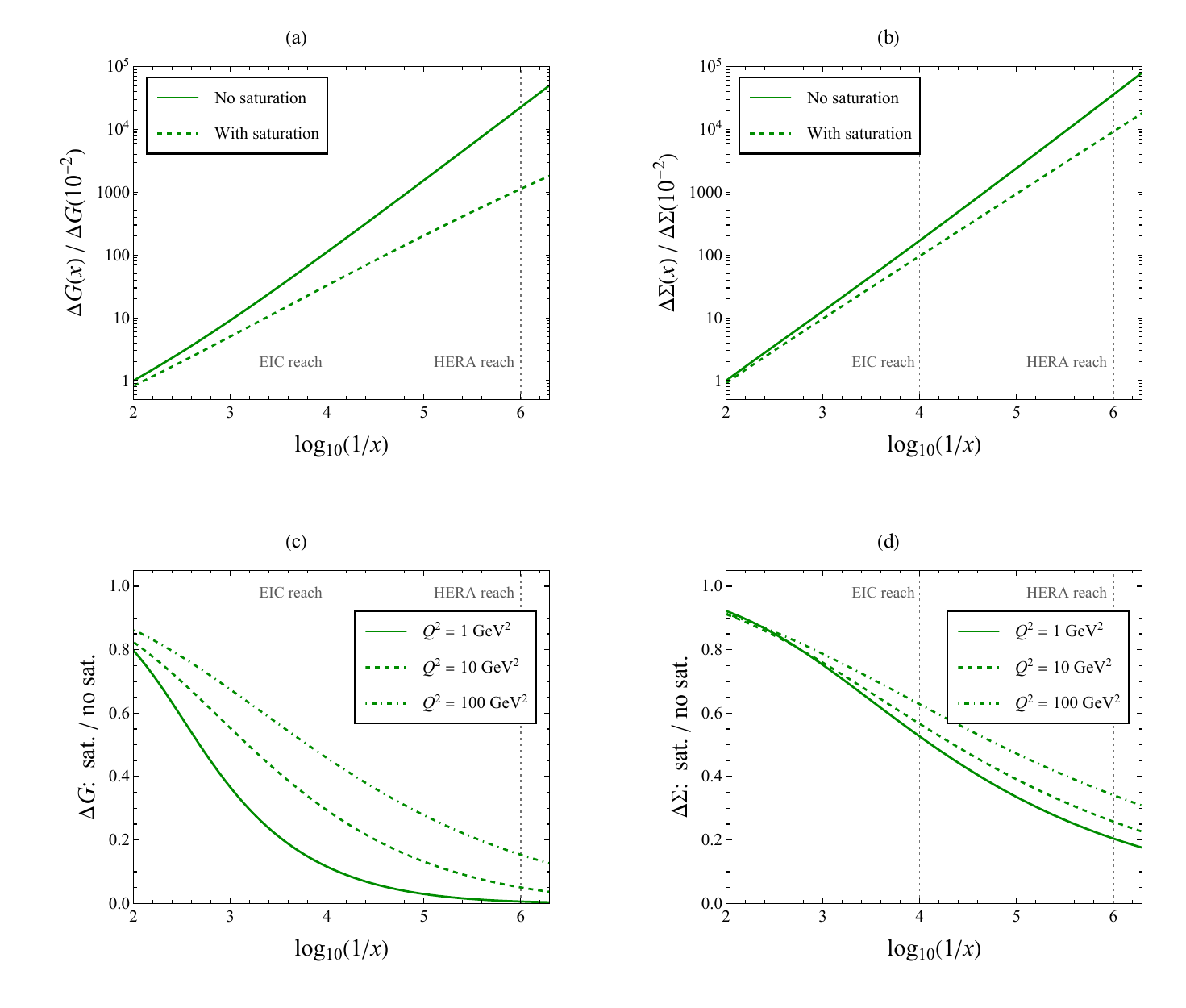}
\caption{Helicity distributions in the experimental range, scheme C, $\alpha_s=0.25$, $\Lambda^2=0.2$~GeV$^2$. (a) $\Delta G$ at $Q^2=10$~GeV$^2$ without saturation (solid) and with saturation (dashed), normalized to the value without saturation at $x=10^{-2}$. (b) The same for $\Delta\Sigma$. (c) Ratio of $\Delta G$ with saturation to $\Delta G$ without saturation at $Q^2=1$ (solid), $10$ (dashed) and $100$~GeV$^2$ (dot-dashed). (d) The same for $\Delta\Sigma$. The vertical lines mark the EIC reach, $x\sim10^{-4}$, and the HERA reach, $x\sim10^{-6}$.}
\label{fig:observables}
\end{figure*}

Figure~\ref{fig:observables} and Tables~\ref{tab:ratios} and \ref{tab:exponents} show $\Delta G$ and $\Delta\Sigma$ in scheme C for $10^{-2}\ge x\ge10^{-6}$. We find the following.

\emph{Saturation suppresses both distributions, $\Delta G$ more strongly.} At $Q^2=10$~GeV$^2$ the ratio of $\Delta G$ with saturation to $\Delta G$ without saturation is $0.82$ at $x=10^{-2}$, $0.29$ at the EIC reach, $x=10^{-4}$, and $0.05$ at the HERA reach, $x=10^{-6}$. For $\Delta\Sigma$ the ratios are $0.91$, $0.57$ and $0.26$. The suppression is stronger at small $Q^2$: at $x=10^{-4}$ it is $0.12$ for $\Delta G$ and $0.53$ for $\Delta\Sigma$ at $Q^2=1$~GeV$^2$, and $0.46$ and $0.63$ at $Q^2=100$~GeV$^2$. It is already present at $x=10^{-2}$, before the target starts to evolve, because the largest dipoles produced by the helicity evolution between $x_0=0.1$ and $x=0.01$ are saturated by the initial condition.

The stronger suppression of $\Delta G$ is the one identified in Sec.~\ref{sec:G2inside}. According to \eq{eq:sysG2}, $G_2$ at the size $1/Q$ is fed by dipoles larger than $1/Q$. These saturate before the dipole $1/Q$ does, so $\Delta G$ feels saturation earlier than the condition $Q_s\Lb x\Rb=Q$ would suggest: in our target $Q_s^2$ stays below $10$~GeV$^2$ in the whole experimental range ($Q_s^2=5.6$~GeV$^2$ at $x=10^{-6}$), while $\Delta G$ at $Q^2=10$~GeV$^2$ is suppressed by a factor of three already at $x=10^{-4}$. The quark helicity integrates over all dipole sizes down to $1/\sqrt{zs}$, most of which are not saturated, and is suppressed less.

\begin{table}[tbp]
\caption{Ratio of the helicity distributions with saturation to those without saturation in scheme C, $\alpha_s=0.25$, $\Lambda^2=0.2$~GeV$^2$. The EIC and HERA reach are $x\sim10^{-4}$ and $x\sim10^{-6}$.}
\label{tab:ratios}
\begin{ruledtabular}
\begin{tabular}{lccccc}
 & $x=10^{-2}$ & $10^{-3}$ & $10^{-4}$ & $10^{-5}$ & $10^{-6}$ \\ \hline
$\Delta G$, $Q^2=1$~GeV$^2$ & 0.80 & 0.37 & 0.12 & 0.029 & 0.006 \\
$\Delta G$, $Q^2=10$~GeV$^2$ & 0.82 & 0.55 & 0.29 & 0.13 & 0.050 \\
$\Delta G$, $Q^2=100$~GeV$^2$ & 0.86 & 0.67 & 0.46 & 0.28 & 0.15 \\
$\Delta\Sigma$, $Q^2=1$~GeV$^2$ & 0.92 & 0.75 & 0.53 & 0.34 & 0.20 \\
$\Delta\Sigma$, $Q^2=10$~GeV$^2$ & 0.91 & 0.76 & 0.57 & 0.39 & 0.26 \\
$\Delta\Sigma$, $Q^2=100$~GeV$^2$ & 0.91 & 0.79 & 0.63 & 0.47 & 0.34
\end{tabular}
\end{ruledtabular}
\end{table}

\emph{Both distributions still grow, with smaller exponents.} In the experimental range neither $\Delta G$ nor $\Delta\Sigma$ reaches a maximum [Fig.~\ref{fig:observables}(a,b)]. We characterize their growth by effective exponents, $O\sim\Lb 1/x\Rb^{\alpha_O}$, fitted in $10^{-4}\ge x\ge10^{-6}$ (Table~\ref{tab:exponents}). For $\alpha_s=0.25$ and $Q^2=10$~GeV$^2$ the exponent of $\Delta G$ is reduced from $1.15$ to $0.77$ and that of $\Delta\Sigma$ from $1.16$ to $0.99$. The reduction is larger for $\Delta G$ and at small $Q^2$: for $Q^2=1$, $10$ and $100$~GeV$^2$ the exponent of $\Delta G$ is $0.50$, $0.77$ and $0.93$, and that of $\Delta\Sigma$ is $0.94$, $0.99$ and $1.05$. Without saturation the exponents are close to, but somewhat below, the asymptotic value $3.66074\sqrt{\bas}$ ($0.98$, $1.26$ and $1.50$ for the three couplings), because the asymptotic regime is not reached in this range. Note that for $\alpha_s=0.25$ the effective exponent of $\Delta\Sigma$ with saturation is close to one. The asymptotic exponents of the equations are discussed in Secs.~\ref{sec:edge} and \ref{sec:convergence}.

\begin{table}[tbp]
\caption{Effective exponents $O\sim\Lb 1/x\Rb^{\alpha_O}$ fitted in $10^{-4}\ge x\ge10^{-6}$, at $Q^2=10$~GeV$^2$, without saturation and with saturation in scheme C. The central values are for $\Lambda^2=0.2$~GeV$^2$, and the ranges correspond to $\Lambda^2=0.05$--$0.8$~GeV$^2$.}
\label{tab:exponents}
\begin{ruledtabular}
\begin{tabular}{lccc}
 & $\alpha_s=0.15$ & $\alpha_s=0.25$ & $\alpha_s=0.35$ \\ \hline
$\Delta\Sigma$, no saturation & 0.88 & 1.16 & 1.39 \\
$\Delta\Sigma$, saturation & 0.75 & 0.99 & 1.19 \\
\quad range in $\Lambda$ & 0.72--0.79 & 0.96--1.03 & 1.15--1.24 \\
$\Delta G$, no saturation & 0.87 & 1.15 & 1.38 \\
$\Delta G$, saturation & 0.54 & 0.77 & 0.96 \\
\quad range in $\Lambda$ & 0.45--0.65 & 0.69--0.87 & 0.87--1.07
\end{tabular}
\end{ruledtabular}
\end{table}

\emph{Dependence on the coupling.} The suppression grows with $\alpha_s$, because the evolution is faster. At $Q^2=10$~GeV$^2$ and $x=10^{-4}$ the ratio for $\Delta G$ is $0.39$, $0.29$ and $0.24$ for $\alpha_s=0.15$, $0.25$ and $0.35$, and that for $\Delta\Sigma$ is $0.70$, $0.57$ and $0.47$.

\subsection{Comparison of the schemes and dependence on the infrared scale}
\label{sec:schemes}

\begin{figure*}[tbp]
\centering
\includegraphics[width=0.9\textwidth]{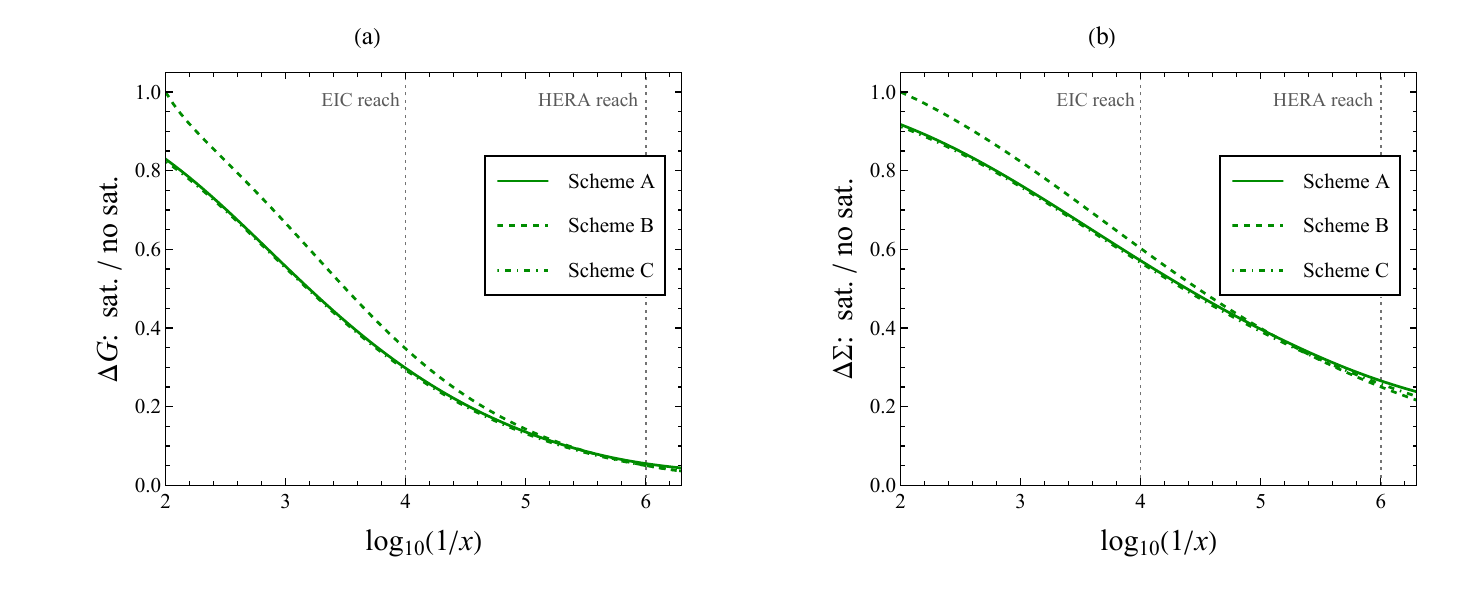}
\caption{Comparison of the three schemes at $Q^2=10$~GeV$^2$, $\alpha_s=0.25$, $\Lambda^2=0.2$~GeV$^2$: ratio with saturation to without saturation for (a) $\Delta G$ and (b) $\Delta\Sigma$, in scheme A (helicity and target from $x_0=0.1$, solid), B (both from $x_0=0.01$, dashed) and C (helicity from $0.1$, target from $0.01$, dot-dashed). The vertical lines mark the EIC and HERA reach.}
\label{fig:schemes}
\end{figure*}

Figure~\ref{fig:schemes} and Table~\ref{tab:schemes} compare the three schemes. Two features stand out.

First, schemes A and C give almost the same results: the ratios differ by at most $3\%$ down to $x=10^{-5}$ and by $10\%$ at $x=10^{-6}$, and the exponents by at most $0.02$. Whether the target is frozen or evolving between $x=0.1$ and $x=0.01$ is therefore immaterial, provided that it reaches the same saturation scale in the experimental range. What matters is where the \emph{helicity} evolution starts. In scheme B, where it starts at $x_0=0.01$, there is no evolution and no suppression at $x=10^{-2}$; the suppression is smaller in the first decades ($0.35$ instead of $0.29$ for $\Delta G$ at the EIC reach), and the exponents are smaller by $0.02$--$0.07$.

Second, the three schemes converge toward small $x$: at the HERA reach the ratios agree within $10\%$. The memory of the starting point of the evolution is lost after about three decades in $x$. This supports the conclusions that do not depend on the first decades of the evolution, in particular the strong suppression of $\Delta G$ and the reduced exponents. At the EIC reach, on the other hand, the choice of $x_0$ is a source of uncertainty of about $20\%$ for $\Delta G$ and $5\%$ for $\Delta\Sigma$.

\begin{table}[tbp]
\caption{Comparison of the three schemes at $Q^2=10$~GeV$^2$, $\alpha_s=0.25$, $\Lambda^2=0.2$~GeV$^2$: ratios with saturation to without saturation, and effective exponents in $10^{-4}\ge x\ge10^{-6}$.}
\label{tab:schemes}
\begin{ruledtabular}
\begin{tabular}{lccccc}
 & $x=10^{-2}$ & $10^{-4}$ & $10^{-6}$ & $\alpha_{\Delta G}$ & $\alpha_{\Delta\Sigma}$ \\ \hline
$\Delta G$, A & 0.83 & 0.30 & 0.055 & 0.78 & \\
$\Delta G$, B & 1 & 0.35 & 0.050 & 0.70 & \\
$\Delta G$, C & 0.82 & 0.29 & 0.050 & 0.77 & \\
$\Delta\Sigma$, A & 0.92 & 0.57 & 0.27 & & 0.99 \\
$\Delta\Sigma$, B & 1 & 0.60 & 0.25 & & 0.95 \\
$\Delta\Sigma$, C & 0.91 & 0.57 & 0.26 & & 0.99
\end{tabular}
\end{ruledtabular}
\end{table}

\emph{Dependence on the infrared scale.} The infrared scale $\Lambda$ fixes which dipoles are saturated at the beginning of the evolution, and it is the largest source of uncertainty (Table~\ref{tab:Lambda}). For $\Lambda^2=0.05$, $0.2$ and $0.8$~GeV$^2$ the ratio for $\Delta G$ at the EIC reach is $0.13$, $0.29$ and $0.51$, and that for $\Delta\Sigma$ is $0.34$, $0.57$ and $0.78$. The exponents depend less on $\Lambda$ (Table~\ref{tab:exponents}). The magnitudes of the suppression are therefore model dependent, while its existence, its larger size for $\Delta G$ than for $\Delta\Sigma$, and its growth toward small $x$ and small $Q^2$ are found for all the parameters that we studied.

\begin{table}[tbp]
\caption{Dependence on the infrared scale $\Lambda$: ratio with saturation to without saturation in scheme C at $Q^2=10$~GeV$^2$ and $\alpha_s=0.25$.}
\label{tab:Lambda}
\begin{ruledtabular}
\begin{tabular}{lcccccc}
 & \multicolumn{3}{c}{$\Delta G$} & \multicolumn{3}{c}{$\Delta\Sigma$} \\
$\Lambda^2$ (GeV$^2$) & $10^{-2}$ & $10^{-4}$ & $10^{-6}$ & $10^{-2}$ & $10^{-4}$ & $10^{-6}$ \\ \hline
0.05 & 0.63 & 0.13 & 0.015 & 0.77 & 0.34 & 0.12 \\
0.2 & 0.82 & 0.29 & 0.050 & 0.91 & 0.57 & 0.26 \\
0.8 & 0.93 & 0.51 & 0.14 & 0.97 & 0.78 & 0.46
\end{tabular}
\end{ruledtabular}
\end{table}

To summarize, in the experimental range saturation reduces the gluon helicity by a factor of three at the EIC reach and by a factor of twenty at the HERA reach ($Q^2=10$~GeV$^2$, scheme C), and the quark helicity by factors of about two and four. Both distributions keep growing in this range, with exponents reduced from $1.15$ to $0.77$ ($\Delta G$) and from $1.16$ to $0.99$ ($\Delta\Sigma$) for $\alpha_s=0.25$. The asymptotic regime of the equations, in which $\Delta G$ reaches a maximum and the exponent of $\Delta\Sigma$ is set by the edge of the saturation region (Sec.~\ref{sec:asympt}), lies outside the experimental range.

\section{Discussion}
\label{sec:discussion}

\emph{Accuracy of the approximation.} Our equations combine the DLA helicity evolution with an unpolarized $S$-matrix that evolves at leading-logarithmic accuracy. As discussed in Sec.~\ref{sec:counting}, this is not justified at strict DLA accuracy, and only a calculation at the next order would settle whether the leading-logarithmic saturation corrections can be kept in this form~\cite{Kovchegov:2015pbl}. Our results should therefore be read as the solution of a well-defined model of helicity evolution with saturation, which reduces exactly to the KPS-CTT evolution in the dilute regime. Two features suggest that this model captures the main physics. First, the structure of the new terms is fixed by the operator structure of Eqs.~(\ref{eq:G_largeNc}) and (\ref{eq:Gi_largeNc}), which is not expected to change at the next order~\cite{Kovchegov:2015pbl}. Second, the asymptotic exponents that we find depend only on the slope $\lambda$ of the saturation boundary, not on the details of $S$ near the boundary. In the experimental range, on the other hand, the magnitudes depend on the infrared region, through $\Lambda$ and through the starting point of the evolution; we have quantified this dependence in Sec.~\ref{sec:schemes}.

\emph{Single-logarithmic terms and running coupling.} We have not included the single-logarithmic terms of the helicity evolution, which were computed in Ref.~\cite{Kovchegov:2021lvz}, nor the running of the coupling in the helicity sector. Both modify the growth of the polarized amplitudes~\cite{Kovchegov:2021lvz,Ermolaev:2003zx}, and it was already noted in Ref.~\cite{Kovchegov:2015pbl} that higher-order corrections could be numerically important for the intercept at realistic couplings. Combining them with saturation would be a natural next step. We expect that the single-logarithmic terms would damp the oscillations of $G_2$ inside the saturation region and turn them into a monotonic decrease, as in the Levin--Tuchin regime of the unpolarized amplitude~\cite{Levin:1999mw,Levin:2001cv} and as found by Li for the sub-eikonal dipole gluon distribution at single-logarithmic accuracy~\cite{Li:2026azt}. This would not change the conclusion that $\Delta G$ stops growing. Our use of a fixed coupling for the helicity evolution together with a running-coupling $S$-matrix is not fully consistent. Beyond leading logarithmic accuracy the evolution of $S$ itself is modified by next-to-leading-order corrections~\cite{Balitsky:2007feb} and by the resummation of collinear logarithms~\cite{Beuf:2014uia,Iancu:2015vea,Ducloue:2019ezk}, which slow down the growth of $Q_s$; the effect of these corrections on the helicity distributions would enter our equations only through $S$, and could be studied with the same method. we have adopted it to keep the helicity sector identical to that of Refs.~\cite{Cougoulic:2022gbk,Borden:2023ugd}, so that every difference can be attributed to saturation.

\emph{Impact parameter.} The equations hold at each impact parameter under the assumption of slow variation with $b$ (Sec.~\ref{sec:counting}), and our numerical results are for a $b$-independent target. For a proton, the saturation scale is smaller at the periphery. Since the asymptotic exponents depend only on the slope $\lambda$ and not on the value of $Q_s$, we expect them to be the same at each $b$ at which the saturation region is eventually reached; in the experimental range, where the magnitudes depend on $Q_s$, the dilute periphery of the proton would reduce the suppression. The magnitudes at moderate $x$, and the possible contribution of a dilute periphery that grows in size with energy, require a $b$-dependent calculation, which we leave for the future.

\emph{Sub-eikonal expansion in strong fields.} The polarized amplitudes are defined through a sub-eikonal expansion of the Wilson lines. Brown, DeBrizzi and Kovner have shown that, for strong classical fields, the naive eikonal expansion mixes eikonal orders, and that the effect of the sub-eikonal components on the eikonal ones is not suppressed by the eikonality parameter~\cite{Brown:2026qlc}. They also proposed a way to rectify this by integrating out the high longitudinal momentum modes. Inside the saturation region the target fields are strong, and the definition of the polarized amplitudes might receive corrections of this type. Our equations do not include them. In a similar spirit, Li showed that in the saturation regime new helicity effects arise from the nonlinear interaction between the eikonal and the sub-eikonal fields~\cite{Li:2024xra}. These effects would enter our equations through the initial conditions and possibly through additional terms in the kernel.

\emph{Large $N_c\,\&\,N_f$.} We have worked at large $N_c$, where the quarks do not take part in the evolution. In the large-$N_c\,\&\,N_f$ limit the evolution includes quark loops and the transition operators of Ref.~\cite{Borden:2024bxa}, and the helicity distributions oscillate in $\ln\Lb 1/x\Rb$~\cite{Kovchegov:2020hgb}. Before the DLA limit is taken, the large-$N_c\,\&\,N_f$ equations of Ref.~\cite{Cougoulic:2022gbk} also contain the unpolarized $S$-matrices of the daughter dipoles, and the method of this paper can be applied to them.

\emph{Relation to earlier work.} Our results are in line with the observation of Itakura \emph{et al.}~\cite{Itakura:2003jp} that saturation reduces the intercept of the DLA Reggeon evolution. In helicity, the reduction is from $3.66\sqrt{\bas}$ to $2\sqrt{1-\lambda}\sqrt{\bas}$ for dipoles inside the saturation region, and to $\phi_0\Lb \lambda\Rb\sqrt{\bas}$ for $\Delta\Sigma$. The mechanism that sets the exponent of $\Delta\Sigma$, the dominance of the dipoles at the edge of the saturation region, is analogous to the role of the saturation boundary in the unpolarized evolution, where the growth of the saturation scale is determined by the linear evolution just outside the saturation region~\cite{Mueller:2002zm,Iancu:2002tr}.

\emph{Phenomenology.} We have not attempted a comparison with data. The analyses of Refs.~\cite{Adamiak:2021ppq,Adamiak:2023yhz} use small-$x$ helicity evolution without saturation, starting at $x_0=0.1$, to extrapolate the helicity PDFs to small $x$. Our results indicate that saturation already matters in the range that the EIC will probe. With a target that reproduces the saturation scale of the proton at HERA, the gluon helicity at $x\sim10^{-4}$ and $Q^2=10$~GeV$^2$ is reduced to $13$--$51\%$ of its value without saturation, and the quark helicity to $34$--$78\%$, the ranges being dominated by the infrared scale $\Lambda$ (Sec.~\ref{sec:schemes}). The suppression grows toward smaller $x$ and smaller $Q^2$, and it would affect the extrapolation of the helicity PDFs beyond the reach of the data. The choice of the starting point of the evolution, $x_0=0.1$ or $0.01$, changes these numbers by about $20\%$ for $\Delta G$ and $5\%$ for $\Delta\Sigma$ at the EIC reach, and its effect disappears toward $x=10^{-6}$. Which choice is more appropriate, and whether the helicity and target evolutions should start at the same $x$, is a question that deserves further discussion; we have presented the three schemes so that the reader can judge. A quantitative statement would require a combined fit, which is beyond the scope of this work.

\section{Conclusions}
\label{sec:conclusions}

We have studied the revised small-$x$ helicity evolution of Ref.~\cite{Cougoulic:2022gbk} at large $N_c$ with the unpolarized dipole $S$-matrices kept in the evolution kernel. Our results are the following.

(i) We derived the double-logarithmic evolution equations~(\ref{eq:system}) for the polarized dipole amplitudes $G$, $G_2$ and their neighbor amplitudes in the presence of saturation. The $S$-matrix of the parent dipole multiplies the emissions in which the polarization goes to a small daughter, the $S$-matrix of the small daughter multiplies the emissions in which it stays in a dipole of the size of the parent, and the $S$-matrix of the large daughters multiplies the IR term of the $G_2$ equation. New terms proportional to $1-S$ appear, among them a term $-\Lb 1-S\Rb\Gamma_2$ in the equation for $G_2$. It arises because the real and virtual ultraviolet contributions, which cancel each other for $S=1$, are weighted differently by the $S$-matrix of the small daughter dipole; by color transparency the new term is ultraviolet finite, and it acts only when the daughter dipole is saturated. For $S=1$ the equations reduce exactly to those of Ref.~\cite{Cougoulic:2022gbk}, and our numerical solution reproduces the intercept $3.66074\sqrt{\bas}$ of Ref.~\cite{Borden:2023ugd} to $0.06\%$.

(ii) Inside the saturation region, where the boundary is the straight line $s=\lambda\eta$, the equations reduce to a hyperbolic equation for the neighbor amplitude, with a coefficient $\pm1$ that depends on whether the daughter is dilute or saturated, and with a boundary condition on the diagonal inherited from the neighbor structure of the KPS-CTT equations. We solved it exactly in the two limits: for $\lambda\to0^+$, $G=I_1\Lb 2U\Rb/U$ with intercept exactly $2\sqrt{\bas}$; for $\lambda=1$, $G=J_1\Lb 2U\Rb/U$ with no growth. For $0<\lambda<1$ a WKB analysis gives $2\sqrt{1-\lambda}\sqrt{\bas}$, which interpolates between the two exact end points and agrees with the numerical solution to better than $1\%$. The amplitude $G_2$ stops growing and oscillates, as $J_1\Lb 2U\Rb/U$ in the limit $\lambda\to1$.

(iii) In the asymptotic regime of the equations, the gluon helicity distribution $\Delta G$ reaches a maximum and decreases, so that the corresponding contribution to the spin sum rule is finite for any coupling. This regime lies outside the experimental range. The quark helicity $\Delta\Sigma$ and the structure function $g_1$ keep growing, with an exponent set by the dipoles at the edge of the saturation region.

(iv) We obtained this exponent in closed form, $\alpha_{\Delta\Sigma}=\phi_0\Lb \lambda\Rb\sqrt{\bas}$, with $\phi_0\Lb c\Rb=\min_{\omega\ge\alpha_h}[\Lb 1-c\Rb\omega+c\,\gamma^-_\omega]$ given by a saddle point of the exact solution of Ref.~\cite{Borden:2023ugd}. The formula agrees with the numerical growth rates without saturation to better than $0.25\%$, and the numerical solution with a saturation boundary of constant slope has its maximal growth rate at the edge of the saturation region, as the formula requires. As a result, the value of $\alpha_s$ below which $\int dx\,\Delta\Sigma$ converges rises from $0.156$ without saturation to $2\pi/[N_c\phi_0^2\Lb \lambda\Rb]$, for example $0.21$ for $\lambda=0.25$ and $0.22$ for $\lambda=0.3$.

(v) In the experimental range $10^{-2}\ge x\ge10^{-6}$, with a running-coupling BK target whose saturation scale grows with $\lambda\simeq0.30$ and is close to that of the proton at HERA, saturation suppresses the helicity distributions without stopping their growth. At $Q^2=10$~GeV$^2$ and $\alpha_s=0.25$, $\Delta G$ is reduced to $29\%$ of its value without saturation at the EIC reach, $x\sim10^{-4}$, and to $5\%$ at $x=10^{-6}$; $\Delta\Sigma$ is reduced to $57\%$ and $26\%$. The effective exponents in $10^{-4}\ge x\ge10^{-6}$ are reduced from $1.15$ to $0.77$ for $\Delta G$ and from $1.16$ to $0.99$ for $\Delta\Sigma$. The magnitudes depend on the infrared scale, by up to a factor of two at the EIC reach. Starting the helicity evolution at $x_0=0.1$, as in the polarized analyses, or at $x_0=0.01$, as in the rcBK descriptions of HERA, changes the results at the EIC reach by about $20\%$ for $\Delta G$, while the evolution of the target between $x=0.1$ and $0.01$ is immaterial; toward $x=10^{-6}$ the three schemes agree.

Saturation thus tames the asymptotic small-$x$ growth of the gluon helicity, and that of the quark helicity only partially, and it already produces a sizable suppression of both distributions in the range of $x$ that the EIC will explore. The natural extensions of this work are the inclusion of the single-logarithmic terms and of the running coupling in the helicity sector, the large-$N_c\,\&\,N_f$ evolution, and a calculation with an impact-parameter dependent target.

\begin{acknowledgments}
I am grateful to Yuri V. Kovchegov, Yossathorn Tawabutr and M. Gabriel Santiago, from whom I learned small-$x$ helicity evolution during my research stay at The Ohio State University in 2021--2022; that experience was the starting point of the present work, which was carried out independently. I thank my colleagues at Universidad de Playa Ancha and Universidad T\'ecnica Federico Santa Mar\'{\i}a for useful discussions, and I acknowledge the use of the computing infrastructure of the Scientific and Technological Center of Valpara\'{\i}so (CCTVal), Universidad T\'ecnica Federico Santa Mar\'{\i}a (ANID PIA/APOYO AFB180002). This work was supported by ANID grant Fondecyt \textnumero 3230699.
\end{acknowledgments}

\section*{Data availability}
The main results of this manuscript are the evolution equations and the analytic expressions derived from them, which are contained within the article. The numerical results shown in the figures and tables are also provided as tabulated plain-text data files in the Supplemental Material~\cite{SupplementalMaterial}.

\appendix

\section{The equations for the neighbor amplitudes}
\label{app:neighbor}

In this appendix we derive Eqs.~(\ref{eq:sysGam}) and (\ref{eq:sysGam2}) for the neighbor amplitudes $\Gamma$ and $\Gamma_2$ in detail, first in transverse coordinates and then in the logarithmic variables~(\ref{eq:vars}).

\emph{Definition.} The neighbor amplitude $\Gamma\Lb x_{10}^2,x_{21}^2,z's\Rb$ has the same operator definition as $G\Lb x_{10}^2,z's\Rb$; it describes a polarized dipole $10$ produced in a step of the evolution in which the adjacent dipole $21$ was also produced~\cite{Kovchegov:2015pbl,Kovchegov:2016zex,Kovchegov:2017lsr,Kovchegov:2018znm,Cougoulic:2022gbk}. The only difference with $G$ is that the subsequent evolution of the dipole $10$ is limited by the lifetime of the neighbor dipole $21$, $z'x_{21}^2$, instead of its own lifetime $z'x_{10}^2$. Since the neighbor amplitudes appear in the UV region, $x_{21}<x_{10}$, the neighbor constraint is the stronger one. The same holds for $\Gamma_2$ and $G_2$.

\emph{The equation for $\Gamma$.} One step of the evolution of $\Gamma\Lb x_{10}^2,x_{21}^2,z's\Rb$ is the emission of a gluon with momentum fraction $z''<z'$ by the dipole $10$, exactly as in Sec.~\ref{sec:Geq}, with a daughter dipole of size $x_{32}$ in the UV region. The bracket is therefore \eq{eq:Gbracket} with $x_{21}\to x_{32}$: the parent dipole is still $10$, so the terms $3G$ and $2G_2$ carry $S\Lb x_{10}^2,z''s\Rb$, while the terms with the neighbor amplitudes carry the $S$-matrix of the small daughter, $S\Lb x_{32}^2,z''s\Rb$. The lifetime ordering of the new emission with respect to the neighbor dipole, $z''x_{32}^2<z'x_{21}^2$, together with $x_{32}<x_{10}$ and $z''x_{32}^2s>1$, gives
\begin{widetext}
\beq
\label{eq:Gam_x}
\begin{split}
\Gamma\Lb x_{10}^2,x_{21}^2,z's\Rb ={}& G^{\Lb 0\Rb}\Lb x_{10}^2,z's\Rb + \frac{\alpha_sN_c}{2\pi}\int\limits_{\frac{1}{sx_{10}^2}}^{z'}\frac{dz''}{z''}\int\limits_{\frac{1}{z''s}}^{\min\left[x_{10}^2,\,x_{21}^2\frac{z'}{z''}\right]}\frac{dx_{32}^2}{x_{32}^2}\,\Big\{\big[2S\Lb x_{32}^2,z''s\Rb-1\big]\Gamma\Lb x_{10}^2,x_{32}^2,z''s\Rb \\
&+ 3S\Lb x_{10}^2,z''s\Rb\,G\Lb x_{32}^2,z''s\Rb + 2S\Lb x_{10}^2,z''s\Rb\,G_2\Lb x_{32}^2,z''s\Rb + 2S\Lb x_{32}^2,z''s\Rb\,\Gamma_2\Lb x_{10}^2,x_{32}^2,z''s\Rb\Big\}.
\end{split}
\eeq
\end{widetext}
For $S=1$ this is the equation for $\Gamma$ of Ref.~\cite{Cougoulic:2022gbk}. We have used $d^2x_{32}/x_{32}^2\to\pi\,dx_{32}^2/x_{32}^2$ after the angular averages of Appendix~\ref{app:UV}.

\emph{The equation for $\Gamma_2$.} The evolution of $\Gamma_2\Lb x_{10}^2,x_{21}^2,z's\Rb$ is the same as that of $G_2$ in Sec.~\ref{sec:G2eq}, with the neighbor lifetime constraint. The IR term (large daughters, $x_{32}>x_{10}$) carries $S\Lb x_{32}^2,z''s\Rb$, and the ordering $z''x_{32}^2<z'x_{21}^2$ now restricts both the size of the daughter and its momentum fraction, $z''<z'x_{21}^2/x_{10}^2$. The UV term is the remnant~(\ref{eq:G2UV}) of the incomplete cancellation of the two UV divergences, with the same limits as in \eq{eq:Gam_x}. Altogether,
\begin{widetext}
\beq
\label{eq:Gam2_x}
\begin{split}
\Gamma_2\Lb x_{10}^2,x_{21}^2,z's\Rb = G_2^{\Lb 0\Rb}\Lb x_{10}^2,z's\Rb &+ \frac{\alpha_sN_c}{\pi}\int\limits_{\frac{\Lambda^2}{s}}^{z'\frac{x_{21}^2}{x_{10}^2}}\frac{dz''}{z''}\int\limits_{\max\left[x_{10}^2,\,\frac{1}{z''s}\right]}^{\min\left[\frac{z'}{z''}x_{21}^2,\,\frac{1}{\Lambda^2}\right]}\frac{dx_{32}^2}{x_{32}^2}\,S\Lb x_{32}^2,z''s\Rb\big[G+2G_2\big]\Lb x_{32}^2,z''s\Rb \\
&- \frac{\alpha_sN_c}{2\pi}\int\limits_{\frac{1}{sx_{10}^2}}^{z'}\frac{dz''}{z''}\int\limits_{\frac{1}{z''s}}^{\min\left[x_{10}^2,\,x_{21}^2\frac{z'}{z''}\right]}\frac{dx_{32}^2}{x_{32}^2}\,\big[1-S\Lb x_{32}^2,z''s\Rb\big]\,\Gamma_2\Lb x_{10}^2,x_{32}^2,z''s\Rb.
\end{split}
\eeq
\end{widetext}
For $S=1$ the second line vanishes and the first line is the equation for $\Gamma_2$ of Ref.~\cite{Cougoulic:2022gbk}.

\emph{Logarithmic variables.} With $\eta''=\sqrt{\bas}\ln\Lb z''s/\Lambda^2\Rb$ and $s_{32}=\sqrt{\bas}\ln\Lb 1/x_{32}^2\Lambda^2\Rb$, the measure becomes $\Lb \alpha_sN_c/2\pi\Rb\,\Lb dz''/z''\Rb\,\Lb dx_{32}^2/x_{32}^2\Rb=d\eta''\,ds_{32}$, and each kinematic condition turns into a linear inequality:
\begin{center}
\begin{tabular}{ll}
\hline\hline
condition & logarithmic form \\ \hline
$z''<z'$ & $\eta''<\eta'$ \\
$z''sx_{10}^2>1$ & $\eta''>s_{10}$ \\
$z''sx_{32}^2>1$ & $s_{32}<\eta''$ \\
$x_{32}<x_{10}$ (UV) & $s_{32}>s_{10}$ \\
$x_{32}>x_{10}$ (IR) & $s_{32}<s_{10}$ \\
$x_{32}<1/\Lambda$ & $s_{32}>0$ \\
$z''x_{32}^2<z'x_{21}^2$ & $s_{32}>s_{21}+\eta''-\eta'$ \\
\hline\hline
\end{tabular}
\end{center}
The last line follows from $\ln\Lb 1/x_{32}^2\Lambda^2\Rb>\ln\Lb 1/x_{21}^2\Lambda^2\Rb-\ln\Lb z'/z''\Rb$. In the UV terms the conditions combine into $s_{10}<\eta''<\eta'$ and $\max[s_{10},\,s_{21}+\eta''-\eta']<s_{32}<\eta''$, which are the limits of \eq{eq:sysGam} and of the second term of \eq{eq:sysGam2}. In the IR term of \eq{eq:Gam2_x} they give $0<s_{32}<s_{10}$ and $s_{32}<\eta''<\eta'-s_{21}+s_{32}$, and the factor $\alpha_sN_c/\pi=2\times\alpha_sN_c/2\pi$ gives the coefficient $2$; this is the first term of \eq{eq:sysGam2}.

\emph{Properties.} (i) On the diagonal, $x_{21}=x_{10}$, the neighbor constraint $z''x_{32}^2<z'x_{10}^2$ is implied by the others, and $\Gamma=G$, $\Gamma_2=G_2$, as used in Sec.~\ref{sec:system}. (ii) The neighbor constraint is what produces the boundary condition~(\ref{eq:bc}) of the reduced problem in Sec.~\ref{sec:asympt}: since the upper limit of the $s_{32}$ integration depends on $s_{21}$ only through $s_{21}+\eta''-\eta'$, the derivative of $\Gamma$ with respect to $s_{21}$ vanishes on the diagonal. (iii) In the numerical solution the neighbor amplitudes are obtained from the recursions of Appendix~\ref{app:numerics}, which are the discrete form of Eqs.~(\ref{eq:Gam_x}) and (\ref{eq:Gam2_x}).

\section{Angular averages of the kernels}
\label{app:UV}

In this appendix we list the angular averages of all the kernels that enter the DLA reduction of Sec.~\ref{sec:DLA}, in the UV region $x_{21}\ll x_{10}$ and in the IR region $x_{21}\gg x_{10}$, and we derive in detail the UV coefficients~(\ref{eq:UVcoef}) of the $G_2$ equation.

\emph{Method.} We write ${\un x}_{10}=\Lb x_{10},0\Rb$, ${\un x}_{21}=r\Lb \cos\varphi,\sin\varphi\Rb$ and ${\un x}_{20}={\un x}_{10}+{\un x}_{21}$, multiply each kernel by $x_{21}^2=r^2$, expand at small $r/x_{10}$ (UV) or at small $x_{10}/r$ (IR), and average over $\varphi$. The resulting number is the coefficient of the logarithmic integral $\int d^2x_{21}/x_{21}^2$; terms that vanish after the angular average, or that fall off faster than $1/x_{21}^2$, do not produce a logarithm. In the DLA the arguments of the amplitudes are evaluated at leading order, $x_{20}\simeq x_{10}$ in the UV region and $x_{20}\simeq x_{21}$ in the IR region; the corrections are derivatives of the amplitudes, which lose one logarithm. All averages were computed analytically and checked numerically.

\emph{Results.} Table~\ref{tab:angular} gives the coefficients in units of $\Lb \alpha_sN_c/2\pi^2\Rb\,d^2x_{21}/x_{21}^2$ (the prefactor $\alpha_sN_c/4\pi^2$ of the second integral of \eq{eq:Gi_largeNc} is included), together with the $S$-matrix that each term carries according to \eq{eq:Sregions}. For the $G$ equation the kernels of $G_2$ and $\Gamma_2$ are those obtained in Ref.~\cite{Cougoulic:2022gbk} after the decomposition~(\ref{eq:decomp}); for the $G_2$ equation we use the kernels of the DLA equation of Ref.~\cite{Cougoulic:2022gbk} before the cancellation of the UV terms.

\begin{table*}[tbp]
\caption{Coefficients of the logarithmic integral $\Lb \alpha_sN_c/2\pi^2\Rb\int d^2x_{21}/x_{21}^2$ produced by each kernel of the large-$N_c$ helicity evolution, in the UV ($x_{21}\ll x_{10}$) and IR ($x_{21}\gg x_{10}$) regions, and the unpolarized $S$-matrix carried by the term. A zero means that the kernel does not produce a double logarithm in that region.}
\label{tab:angular}
\begin{ruledtabular}
\begin{tabular}{llccl}
Equation & Kernel and amplitude & UV & IR & $S$-matrix \\ \hline
$G$ & $2\big[\frac{1}{x_{21}^2}-\frac{{\un x}_{21}\cdot{\un x}_{20}}{x_{21}^2x_{20}^2}\big]\times G_{21}$ & 2 & 0 & $S_{20}\to S\Lb x_{10}^2\Rb$ \\
$G$ & same kernel $\times\,\Gamma_{20,21}$ & 2 & 0 & $S_{21}=S\Lb x_{21}^2\Rb$ \\
$G$ & $\frac{x_{10}^2}{x_{21}^2x_{20}^2}\times G_{12}$ & 1 & 0 & $S_{20}\to S\Lb x_{10}^2\Rb$ \\
$G$ & same kernel $\times\,\Lb -\Gamma_{10,21}\Rb$ & $-1$ & 0 & none (virtual) \\
$G$ & kernel of $G_2\Lb x_{21}^2\Rb$ & 2 & 0 & $S_{20}\to S\Lb x_{10}^2\Rb$ \\
$G$ & kernel of $\Gamma_2\Lb x_{20}^2,x_{21}^2\Rb$ & 2 & 0 & $S_{21}=S\Lb x_{21}^2\Rb$ \\ \hline
$G_2$ & virtual, $-\frac{{\un x}_{10}\cdot{\un x}_{21}}{x_{21}^2x_{20}^2}\times G_2\Lb x_{21}^2\Rb$ & 0 & 0 & none \\
$G_2$ & virtual, $-\frac{x_{10}^2}{x_{21}^2x_{20}^2}\times\Gamma_2\Lb x_{10}^2,x_{21}^2\Rb$ & $-1$ & 0 & none \\
$G_2$ & real, kernel of $G\Lb x_{21}^2\Rb$ & 0 & 1 & $S\Lb x_{21}^2\Rb$ \\
$G_2$ & real, kernel of $\Gamma\Lb x_{20}^2,x_{21}^2\Rb\to G\Lb x_{21}^2\Rb$ & 0 & 1 & $S\Lb x_{21}^2\Rb$ \\
$G_2$ & real, kernel of $G_2\Lb x_{21}^2\Rb$ & 0 & 2 & $S\Lb x_{21}^2\Rb$ \\
$G_2$ & real, $\mathcal{M}\times\Gamma_2\Lb x_{20}^2,x_{21}^2\Rb$ & $+1$ & 2 & $S\Lb x_{21}^2\Rb$
\end{tabular}
\end{ruledtabular}
\end{table*}

\emph{Reading the table.} In the $G$ equation only the UV region contributes. The first four lines give $2S\Lb x_{10}^2\Rb G+2S\Lb x_{21}^2\Rb\Gamma+S\Lb x_{10}^2\Rb G-\Gamma$, which is \eq{eq:Gbracket1}, and the next two give \eq{eq:Gbracket2}. For $S=1$ the coefficients $1$ (of $\Gamma$), $3$ (of $G$), $2$ (of $G_2$) and $2$ (of $\Gamma_2$) of Ref.~\cite{Cougoulic:2022gbk} are recovered. In the $G_2$ equation the IR region gives $1+1=2$ for $G$ and $2+2=4$ for $G_2$, all multiplied by $S\Lb x_{21}^2\Rb$, which is \eq{eq:G2IR}; for $S=1$ this is the combination $2[G+2G_2]$ of Ref.~\cite{Cougoulic:2022gbk}. In the UV region of the $G_2$ equation only the pair of $\Gamma_2$ terms survives, with coefficients $-1$ and $+1\times S\Lb x_{21}^2\Rb$, which is \eq{eq:UVcoef}. We now give the details of this last calculation.

\emph{The UV pair.}

After contraction with $\epsilon^{ij}x_{10}^j$ and integration over the impact parameter, the two terms of \eq{eq:Gi_largeNc} that multiply the neighbor amplitude $\Gamma_2\Lb x_{10}^2,x_{21}^2\Rb$ in the region $x_{21}\ll x_{10}$ are, in the DLA equation for $G_2$ of Ref.~\cite{Cougoulic:2022gbk},
\beq
\label{eq:UVterms}
\begin{split}
&-\frac{\alpha_sN_c}{2\pi^2}\int\frac{dz'}{z'}\int d^2x_2\,\frac{x_{10}^2}{x_{21}^2x_{20}^2}\,\Gamma_2 \\
&+ \frac{\alpha_sN_c}{4\pi^2}\int\frac{dz'}{z'}\int\frac{d^2x_2}{x_{10}^2}\,\mathcal{M}\Lb {\un x}_{10},{\un x}_{21}\Rb\,S_{21}\,\Gamma_2,
\end{split}
\eeq
where the first term is the virtual one and
\beq
\label{eq:Mkernel}
\begin{split}
\mathcal{M} ={}& {\un x}_{10}\cdot{\un x}_{20}\left(\frac{3}{x_{21}^2}-2\frac{{\un x}_{20}\cdot{\un x}_{21}}{x_{20}^2x_{21}^2}-\frac{1}{x_{20}^2}\right) \\
&+ 2\frac{\Lb {\un x}_{21}\times{\un x}_{20}\Rb^2}{x_{21}^2}\left(-\frac{1}{x_{21}^2}+2\frac{{\un x}_{20}\cdot{\un x}_{21}}{x_{20}^2x_{21}^2}+\frac{1}{x_{20}^2}\right)
\end{split}
\eeq
is the kernel of the real term~\cite{Cougoulic:2022gbk}, in which we have restored the factor $S_{21}$ according to \eq{eq:Gi_largeNc}. We write ${\un x}_{21}=r\Lb \cos\varphi,\sin\varphi\Rb$, ${\un x}_{20}={\un x}_{10}+{\un x}_{21}$, and expand at small $r$. Both kernels behave as $1/r^2$, and their angular averages are
\beq
\label{eq:angav}
\Big\langle r^2\,\frac{x_{10}^2}{x_{21}^2x_{20}^2}\Big\rangle_\varphi \to 1, \qquad \Big\langle r^2\,\frac{\mathcal{M}}{x_{10}^2}\Big\rangle_\varphi \to 2 .
\eeq
We have obtained both limits analytically, expanding the integrands in $r$ and integrating over $\varphi$, and checked them numerically for $r/x_{10}=10^{-2}$, $10^{-3}$ and $10^{-4}$. With the prefactors of \eq{eq:UVterms}, the virtual term contributes $-1$ and the real term $+1\times S_{21}$, in units of $\Lb \alpha_sN_c/2\pi^2\Rb\,d^2x_{21}/x_{21}^2$, which is \eq{eq:UVcoef}. For $S_{21}=1$ the two cancel, as stated in Ref.~\cite{Cougoulic:2022gbk}, and for $S_{21}\neq1$ they leave the term~(\ref{eq:G2UV}).

\section{Numerical method}
\label{app:numerics}

\emph{Discretization.} We discretize the $\Lb s,\eta\Rb$ plane with a step $\delta$, $s_{10}=i\delta$, $\eta=j\delta$, $s_{21}=k\delta$, and write $G_{ij}=G\Lb i\delta,j\delta\Rb$, $\Gamma_{ikj}=\Gamma\Lb i\delta,k\delta,j\delta\Rb$, and similarly for $G_2$, $\Gamma_2$ and $S_{ij}=S\Lb i\delta,j\delta\Rb$, as in Ref.~\cite{Cougoulic:2022gbk}. The integrals in \eq{eq:system} become sums. It is convenient to work with the differences between consecutive rapidities, which turns the equations into recursions. Defining, at the rapidity $j-1$,
\beq
\label{eq:Aij}
\begin{split}
A_{ii'} ={}& \Gamma_{ii'} + 3S_{i}\,G_{i'} + 2S_{i}\,G_{2,i'} + 2\Gamma_{2,ii'} \\
&- 2\Lb 1-S_{i'}\Rb\big(\Gamma_{ii'}+\Gamma_{2,ii'}\big), \\
W_{ii'} ={}& \Lb 1-S_{i'}\Rb\,\Gamma_{2,ii'},
\end{split}
\eeq
where all quantities are at rapidity index $j-1$ and $i\le i'\le j-1$, we have
\begin{subequations}
\label{eq:recursions}
\begin{align}
G_{ij} &= G_{i,j-1} + \Delta G^{\Lb 0\Rb} + \delta^2\sum_{i'=i}^{j-1}A_{ii'}, \\
\Gamma_{ikj} &= \Gamma_{i,k-1,j-1} + \Delta G^{\Lb 0\Rb} + \delta^2\sum_{i'=k-1}^{j-1}A_{ii'}, \quad i<k, \\
G_{2,ij} &= G_{2,i,j-1} + 2\delta^2\sum_{i'=0}^{i-1}S_{i',i'+j-i}\big[G+2G_2\big]_{i',i'+j-i} \notag \\
&\quad - \delta^2\sum_{i'=i}^{j-1}W_{ii'}, \\
\Gamma_{2,ikj} &= \Gamma_{2,i,k-1,j-1} - \delta^2\sum_{i'=k-1}^{j-1}W_{ii'}, \quad i<k,
\end{align}
\end{subequations}
with $\Gamma_{iij}=G_{ij}$, $\Gamma_{2,iij}=G_{2,ij}$, $\Delta G^{\Lb 0\Rb}=\mathcal{N}\Lb C_F-2\Rb\delta$, and the Born values~(\ref{eq:Born_vars}) for $j\le i$. For $S=1$ these recursions reduce to the discretized equations of Ref.~\cite{Cougoulic:2022gbk}. In that case the recursion for $\Gamma_2$ can be iterated down to the diagonal, which gives the closed form $\Gamma_{2,ikj}=G_{2,i,j-k+i}$ [the inhomogeneous term of $G_2$ does not depend on $\eta$]. We use this closed form in the first of our two solutions, which treats the equations of Ref.~\cite{Cougoulic:2022gbk} only. The second solves Eqs.~(\ref{eq:recursions}) with a general $S$. For $S=1$ the two agree on the whole grid within round-off.

\emph{Convergence.} The growth rates inside the saturation region with the step boundary change by $0.6\%$ ($\lambda=0.25$) and $8\%$ ($\lambda=0.5$) between $\delta=0.1$ and $\delta=0.05$ at $\eta=25$--$40$. The larger change for large $\lambda$ comes from the discretization of the boundary $s=\lambda\eta$, which crosses more grid cells. Replacing the step by the GBW form~(\ref{eq:GBW}) changes the rates by $1\%$ for $\lambda=0.25$ and by up to $20\%$ for $\lambda=0.5$ at $s_{10}=2$, because the width $\sqrt{\bas}$ of the GBW boundary is not small compared with the distances in $s$ that matter for large $\lambda$. In the experimental range (scheme C, $\alpha_s=0.25$, $\Lambda^2=0.2$~GeV$^2$, $Q^2=10$~GeV$^2$) the ratios with saturation to without saturation change by less than $3\%$ for $x\ge10^{-4}$ and by $6\%$ at $x=10^{-6}$ for $\Delta G$ between $\delta=0.1$, $0.05$ and $0.025$, and the effective exponents by less than $1\%$ (for $\Delta\Sigma$: $0.990$, $0.990$ and $0.991$; for $\Delta G$: $0.768$, $0.770$ and $0.776$).

\emph{Reduced equations.} Equations~(\ref{eq:redGam_uv}) and (\ref{eq:G2red}) are solved with the same recursion as the full system, \eq{eq:recursions}, keeping only the neighbor amplitude of a parent at $s_{10}=0$ and replacing the bracket by $\sigma\,\Gamma$ (or $\sigma_2\Gamma_2$, with $\sigma_2=-1$ for saturated daughters and $0$ otherwise). This preserves the triangular integration region and hence the diagonal condition~(\ref{eq:bc}). For $\sigma=\pm1$ the results, extrapolated linearly to $\delta\to0$ from $\delta=0.02$ and $0.01$ (or $0.01$ and $0.005$), reproduce the exact solutions~(\ref{eq:Gexact0}) and (\ref{eq:Gexact1}). The growth rates in Table~\ref{tab:omega} are slopes of the upper envelope of $\ln|G|$, obtained from the maxima of $|G|$ in consecutive windows of length $\Delta\eta=3$.

\emph{Observables and exponents.} $\Delta G$ is computed from \eq{eq:DG_DLA} at $s_Q=\sqrt{\bas}\ln\Lb Q^2/\Lambda^2\Rb$, rounded to the grid, and $\Delta\Sigma$ from \eq{eq:DSigma_vars} by a double sum; the value of $x$ follows from $y=\sqrt{\bas}\ln\Lb x_0/x\Rb$. Effective exponents are slopes of linear fits of $\ln|O|$ versus $\ln\Lb 1/x\Rb$ in $10^{-4}\ge x\ge10^{-6}$. The growth rates along rays are slopes of $\ln|G\Lb c\,\eta,\eta\Rb|$ in $\eta$, with $s=c\,\eta$ rounded to the grid.

\emph{rcBK.} Equation~(\ref{eq:rcBK}) is solved on a logarithmic grid of $260$ points in $\ln r\in[-22,5]$ ($r$ in GeV$^{-1}$), with $48$ angles in $\Lb 0,\pi\Rb$ for the integration over ${\un r}_1$, linear interpolation of $N\Lb r_2\Rb$ in $\ln r_2$ (with $N=0$ and $N=1$ outside the grid at small and large $r$, respectively), and a fourth-order integration in $Y$ with step $\Delta Y=0.2$. The solution is stored every $\Delta Y=0.2$ and interpolated bilinearly in $\Lb \ln r,Y\Rb$ when \eq{eq:Smap} is evaluated. The saturation scale is obtained by linear interpolation of $N$ between the points of the grid in $\ln r$. The initial saturation scale of scheme A, $Q_{s0}^2=0.036$~GeV$^2$, was obtained by minimizing the root-mean-square difference of $\ln Q_s^2$ between scheme A and schemes B and C in $10^{-2}\ge x\ge10^{-6}$; the remaining difference is below $4\%$.

\section{Small-$c$ expansion of the saddle point}
\label{app:saddle}

Near the branch point the inner square root of \eq{eq:gammaminus} vanishes linearly. Writing
\beq
\label{eq:Dw}
D\Lb \omega\Rb = 1-\frac{16}{\omega^2}\sqrt{1-\frac{4}{\omega^2}}, \qquad D\Lb \alpha_h\Rb=0,
\eeq
and $\omega=\alpha_h+\epsilon$, one has $D\simeq D'\Lb \alpha_h\Rb\,\epsilon$ and
\beq
\label{eq:gm_exp}
\gamma^-_\omega \simeq \frac{\alpha_h}{2} + \frac{\epsilon}{2} - A\sqrt{\epsilon}, \qquad A = \frac{\alpha_h}{2}\sqrt{D'\Lb \alpha_h\Rb} .
\eeq
Numerically $D'\Lb \alpha_h\Rb=0.43011$ and $A=1.20041$. Inserting \eq{eq:gm_exp} into $f_c\Lb \omega\Rb$ of \eq{eq:omegaint},
\beq
f_c = \alpha_h\Big(1-\frac{c}{2}\Big) + \Big(1-\frac{c}{2}\Big)\epsilon - cA\sqrt{\epsilon},
\eeq
which is minimal at $\sqrt{\epsilon^\ast}=cA/[2\Lb 1-c/2\Rb]$. Hence
\begin{align}
\label{eq:phi0_small}
\phi_0\Lb c\Rb &\simeq \alpha_h\Big(1-\frac{c}{2}\Big) - \frac{A^2c^2}{4\,\Lb 1-c/2\Rb}, \\
\label{eq:wstar_small}
\omega^\ast\Lb c\Rb &\simeq \alpha_h + \frac{A^2c^2}{4\,\Lb 1-c/2\Rb^2}. 
\end{align}
The first term of \eq{eq:phi0_small} is the branch-point estimate of Sec.~\ref{sec:edge}; the second term is the correction due to the saddle point, which is always negative. Compared with the exact minimization~(\ref{eq:phi0}), \eq{eq:phi0_small} is accurate to $4\times10^{-10}$ for $c=0.01$, $5\times10^{-6}$ for $c=0.1$ and $9\times10^{-5}$ for $c=0.2$, and \eq{eq:wstar_small} reproduces $\omega^\ast$ to $3\times10^{-4}$ for $c\le0.2$. Since the saddle point moves away from the branch point only quadratically in $c$, the branch point controls the intercept at $c=0$. For the physically relevant edge rays, $c=\lambda\approx0.2$--$0.3$, the branch-point estimate overestimates $\phi_0$ by $0.5$--$1.3\%$, and the difference grows quickly for larger $c$ [Fig.~\ref{fig:edge}(b)].

\end{document}